%% file: main.tex
\pdfoutput=1
\documentclass[sigplan,10pt]{acmart}

\input{preamble-common}

\input{preamble-acm-preprint}
\input{macros}

\input{frontmatter-arxiv}

\begin{document}

\begin{abstract}
\input{abstract-arxiv}
\end{abstract}
\input{sections/keywords}
\maketitle

\input{sections/01-introduction}
\input{sections/02-background}
\input{sections/03-configuration}
\input{sections/04-memory}
\input{sections/05-time_breakdown}
\input{sections/07-systems_measure}
\input{sections/08-quality}
\input{sections/09-related}
\input{sections/10-discussion}
\input{sections/11-conclusion}

\bibliographystyle{ACM-Reference-Format}
\bibliography{references}

\end{document}

%% file: preamble-common.tex
\usepackage[utf8]{inputenc}

\usepackage{amsmath}
\usepackage{booktabs}       %
\usepackage{graphicx}
\usepackage{xcolor}

\usepackage{tikz}
\usetikzlibrary{
  arrows.meta,   %
  positioning,   %
  calc,          %
  fit,           %
  backgrounds,   %
  decorations.pathreplacing,  %
}
\usepackage{microtype}      %
\usepackage{subcaption}
\usepackage{multirow}
\usepackage{siunitx}        %
\usepackage{xspace}         %
\usepackage{balance}        %

\usepackage{listings}
\AtBeginDocument{%
}

%% file: preamble-acm-preprint.tex
\setcopyright{none}
\renewcommand\footnotetextcopyrightpermission[1]{}

\newcommand{\figsmall}{\columnwidth}

\newenvironment{widefigure}{\begin{figure*}}{\end{figure*}}
\newenvironment{widetable}{\begin{table*}}{\end{table*}}

\AtBeginDocument{\pagestyle{plain}}

%% file: macros.tex
\def\FINAL{1}
\newif\ifDRAFT
\ifx\FINAL\undefined \DRAFTtrue \else \DRAFTfalse \fi

\ifDRAFT
  \newcommand{\todo}[1]{\textcolor{red}{\textbf{[TODO: #1]}}}
  \newcommand{\note}[2]{\textcolor{blue}{\textbf{[#1: #2]}}}
  \newcommand{\cut}[1]{\textcolor{gray}{#1}}
\else
  \newcommand{\todo}[1]{}
  \newcommand{\note}[2]{}
  \newcommand{\cut}[1]{}
\fi

\newcommand{\smollm}{\mbox{SmolLM3-3B}\xspace} %
\newcommand{\lamp}{\mbox{LaMP}\xspace} %
\newcommand{\rougeone}{\mbox{ROUGE-1}} %
\newcommand{\mae}{\textsc{MAE}} %
\newcommand{\acc}{\textsc{Acc}} %

\newcommand{\best}[1]{\bfseries #1} %
\newcommand{\up}{$\uparrow$} %
\newcommand{\down}{$\downarrow$} %

%% file: frontmatter-arxiv.tex
\title{Fine-Tuning a 3B-Parameter LLM on a Smartphone: Characterizing Sustained Training}

\author{Andrew Geyko}
\affiliation{%
  \institution{Saarland University}
  \city{Saarbr\"ucken}
  \country{Germany}
}
\email{ange00008@uni-saarland.de}

\author{Marius Mosbach}
\affiliation{%
  \institution{Saarland University}
  \city{Saarbr\"ucken}
  \country{Germany}
}
\email{marius.mosbach@uni-saarland.de}

\author{Andr\'e Brinkmann}
\affiliation{%
  \institution{Saarland University}
  \city{Saarbr\"ucken}
  \country{Germany}
}
\email{andre.brinkmann@uni-saarland.de}

%% file: abstract-arxiv.tex
Multi-billion-parameter LLMs now run on phones for inference, and training them on the device would personalize them without user data leaving the phone.
Prior work has measured individual training steps of such models on phones, but not complete training runs, and not whether adapters trained on the device improve personalization.
We present the first systematic characterization of a multi-billion-parameter LLM fine-tuned on a mobile device, covering memory, per-step time, thermal behavior, and energy.
An iPhone~17~Pro can fine-tune a 3B-parameter LLM to a typical user within one battery charge, and the resulting adapters improve personalization as much as adapters trained on a server.
Sustained training throttles the phone to about half its initial throughput, and none of the pausing or burst schedules we tested recovers it.
Nearly all of each training step is spent in the frozen base model, most of it in the backward pass, which nine of the ten other runtimes we audited do not accelerate.
Apple's MLX had a kernel for it that was never dispatched and was incorrect, and our repair, now merged upstream, trains an adapter 1.47$\times$ faster on a third less energy.
On-device fine-tuning is feasible on current phones, and making it efficient requires runtimes and operating systems to treat training as a first-class workload.

%% file: sections/keywords.tex
\ccsdesc[500]{Computing methodologies~Machine learning}
\ccsdesc[300]{Computer systems organization~Embedded systems}

\keywords{on-device training, LoRA, quantization, mobile GPU, thermal throttling, energy, measurement study}

%% file: sections/01-introduction.tex
\section{Introduction}
\label{sec:intro}

\begin{figure}[t]
    \centering
    \includegraphics[width=\linewidth]{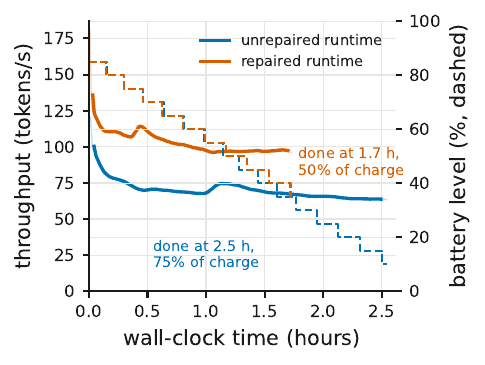}
    \caption{One complete per-user adapter run on an iPhone~17~Pro under each backward kernel.}
    \label{fig:hook}
\end{figure}

LLMs are now a standard component of mobile operating systems, and both iOS and Android ship an on-device model~\cite{applefm2025,gemini2023}.
The same phones hold their users' messages and emails, from which a model could learn, for example, to draft replies in the user's own style~\cite{zhang2025personalizationlargelanguagemodels}.
However, the models deployed on phones today do not learn from this data.
They are trained once on public data and ship with identical weights to every user, so any personalization is limited to the user data that prompt retrieval can place in context at the time of a request~\cite{lamp2024,salemi2024optimizingpersonalization}.
Fine-tuning on the history instead encodes the user's patterns in the model's weights without lengthening the prompt, with a benefit that grows with the length of the history and adds to that of prompt retrieval~\cite{oppu2024,salemi2024ragvsft}.
Training has traditionally run on servers, but user history is sensitive, and even the gradients and activations that federated and split learning send off the device can be used to reconstruct it~\cite{zhu2019dlg,he2019modelinversion,pasquini2021unleashtiger}.
In contrast, training on the device keeps the history, and everything computed from it, on the phone.
We present the first systematic characterization of a multi-billion-parameter LLM fine-tuned on a mobile device.

Fine-tuning a model on one user's history is small by server standards, a few thousand training steps over a few hundred records.
Full fine-tuning keeps a gradient and two optimizer values for every parameter and needs several times the memory of inference, but three techniques from server-side training bring this training within a phone's memory: low-rank adaptation (LoRA)~\cite{hu2021loralowrankadaptationlarge} trains a small adapter while the base model stays frozen, quantization~\cite{dettmers2023qlora} stores weights in a lower precision, and gradient checkpointing~\cite{chen2016gradientcheckpointing} recomputes activations instead of storing them.
However, sustained training of models with billions of parameters has not yet been studied on a phone.
Prior work has measured only individual training steps of such models on a phone~\cite{mobilemebpsong25,park2026mesp,fblayout_tam26}, and has fine-tuned models of only about 1B parameters~\cite{mobilefinetuner2025,wang2025xpert,wang2026edgetune}, split larger models across several mobile devices~\cite{chen2025confidant}, or trained without backpropagation~\cite{pocketllm2024published,gao2025mobizo}.
A history of a few hundred records takes hours to train, and over that time the phone throttles as it heats and drains its battery.
Furthermore, models are trained in runtimes and operating systems designed for inference, where each generated token takes milliseconds and each weight matrix is read in a single direction.

We measure per-user training of a 3B-parameter model at 4-bit precision on an iPhone~17~Pro, whose GPU cores include matrix units~\cite{applea19pro}.
The measurements run from the GPU kernels of one training step to the battery drain of a complete run and cover memory, the time decomposition of a step, sustained thermal behavior, and energy.
We complement these measurements with an audit of the quantized backward pass in ten other runtimes and an evaluation of the resulting adapters on a personalization benchmark.

\autoref{fig:hook} shows two complete runs of the same adapter, one before and one after a kernel repair described below.
In both, throughput falls within minutes as the phone throttles and stays low for the hours the run takes.
None of the pausing or burst schedules we tested recovers the lost throughput.
Over complete runs, training time and energy are linear in the number of training tokens, about 0.01 seconds and 0.04 joules per token, so the cost of an adapter can be predicted before a run starts.
An adapter over a 405-record history costs about half a battery charge, and a 987-record history exhausts a full charge before training completes.
Adapters trained on the phone improve personalization scores on top of prompt retrieval and match adapters trained on server GPUs.

Although adapters are small, every training step still runs the input through the frozen model and propagates gradients back through it~\cite{chen2025celora,sung2022lst,tenison2026lars}.
On the phone, the frozen model takes nearly all of each step, most of it in the backward pass, which reads the quantized weights in the opposite direction from inference.
MLX~\cite{mlx2023}, the runtime we use, contained a matrix-unit kernel for this direction that had never been dispatched and was incorrect.
The repaired kernel trains an adapter 1.47$\times$ faster, on a third less energy, with an unchanged loss trajectory, and every other end-to-end measurement in this paper uses it.
Because runtimes are designed for inference, only one of the ten other runtimes we audited, ExecuTorch~\cite{executorch2026,executorchq4gsw}, provides an accelerated kernel for this direction, added concurrently with this work.

This paper makes the following contributions.
\begin{itemize}
  \item The first measurements of complete multi-billion-parameter training runs on a phone, covering training time, sustained throttling, adapter configuration, and energy (\autoref{sec:sustained}), and the memory and per-step breakdowns they rest on (\autoref{sec:memory}, \autoref{sec:time}).
  \item A diagnosis of the quantized backward pass that exposed an incorrect, never-dispatched kernel in MLX, and an audit showing that only one of ten other runtimes provides an accelerated kernel for this path (\autoref{sec:time}).
  \item An evaluation showing that adapters trained on the phone improve personalization benchmark scores on top of prompt retrieval and match adapters trained on server GPUs (\autoref{sec:quality}).
  \item Recommendations for runtimes and operating systems that support on-device training (\autoref{sec:discussion}).
\end{itemize}
\anon[]{Our code and measurement data are available at \url{https://github.com/gordofreemo/mobile_LoRA_ft}.}

\autoref{sec:background} introduces personalization by fine-tuning, the techniques that fit training on a phone, and the iOS execution environment, and \autoref{sec:setup} describes the experimental setup.
\autoref{sec:memory} measures memory, \autoref{sec:time} the time within a training step and the kernel repair, and \autoref{sec:sustained} complete adapter runs.
\autoref{sec:quality} evaluates the adapters, \autoref{sec:related} discusses related work, \autoref{sec:discussion} discusses implications and limitations, and \autoref{sec:conclusion} concludes.

%% file: sections/02-background.tex
\section{Background}
\label{sec:background}
We describe how fine-tuning personalizes an LLM, the techniques that bring this training within a phone's memory, and the constraints iOS places on an application that trains, since these determine the cost of the workload we measure.

\subsection{LLM Fine-Tuning for Personalization}
\label{sec:ft-personalization}
Personalization adapts a model to a user based on that user's past behavior.
\lamp{}~\cite{lamp2024} is an established benchmark to measure the ability of a model to personalize its responses to users.
It consists of seven prediction tasks, each pairing a query with the user's history of similar items, which differentiates users from one another.
For example, the rating task provides a user review and the user's earlier reviews with ratings, and asks the model to predict what rating the user would give a product.

A model can use the history in two ways.
Retrieval selects the few past records most similar to the new item and places them in context.
Fine-tuning instead trains on the history, so that the user's patterns are encoded in the model's weights.
Retrieval requires no training and adapts to new records at inference time.
Fine-tuning requires training but can draw on the entire history and is not limited by the model's context length. Prior work has found that the benefit of fine-tuning grows with the length of a user's history and adds to that of retrieval~\cite{oppu2024,salemi2024ragvsft}.

Personalized fine-tuning demands two things from a model.
The model must first learn the task, which requires a diverse training set from many users, and it must learn about the individual, which requires only the records from that user.
\citet{oppu2024} separate the two into successive stages, each trained as an adapter, a small set of additional weights that is trained while the base model stays frozen and that can be added into the base weights afterwards.
A \textbf{task adapter} is trained on a corpus drawn from many users, and is then merged into the base weights so that the result behaves as an ordinary model that knows the task.
A \textbf{per-user adapter} is then trained on top of that model from a single user's history, with each past record serving as one training example.
The task stage needs data that no single device holds and can run once on a server.
The per-user stage needs only the data the device already has.
This division makes on-device fine-tuning feasible, and the per-user stage is the run this paper measures.

\subsection{On-Device Training}
\label{sec:ondevice_training}
Training requires more memory than inference.
The activations of every block stay resident until the backward pass consumes them, and every trained parameter carries a gradient and two optimizer values, three extra values per parameter.
For a model of several billion parameters, full fine-tuning at 16-bit precision therefore requires several times the memory of inference.
In order to fit training on-device, three techniques adopted from server-side training are required.

\paragraph{Low-rank Adaptation.}
LoRA~\cite{hu2021loralowrankadaptationlarge} reduces the number of trained parameters by keeping every weight matrix $W$ frozen and adding a smaller weight matrix $BA$ to it, where $B$ and $A$ are thin matrices whose shared inner dimension, the rank, is small.
Only $B$ and $A$ receive gradient and optimizer states, whose parameter counts amount to a fraction of a percent of the original model.
Every training step still does a full forward pass through the base model, and a full backward pass still runs through $W$ to carry the gradient to previous blocks. Only the gradient with respect to $W$ itself disappears.
LoRA therefore removes nearly all gradient and optimizer memory but only about a third of the arithmetic, the products that compute the gradient of each $W$~\cite{hu2021loralowrankadaptationlarge,chen2025celora}.

\paragraph{Quantization.}
Quantization reduces the precision of every parameter, lowering the size of the base model.
Each weight is stored as a low-precision integer together with a scale and offset that reconstruct the value.
The scale and offset are shared by a group of $G$ consecutive weights along the input dimension of the layer, where $G$ is the group size.
At 4 bits with groups of 64, the model occupies roughly a fourth of the original memory footprint of the base 16-bit model.
The grouping also fixes how weights are laid out in memory, a fact that becomes important when the backward pass reads those weights in a different order (\autoref{sec:time:backward_pass}).

\paragraph{Gradient Checkpointing.}
Checkpointing reduces the activation memory at the cost of recomputation.
Instead of storing every block's activations through the forward pass, the training loop stores only the inputs of selected blocks and recomputes the rest during the backward pass~\cite{chen2016gradientcheckpointing}.
The selection can be made per block or per group of blocks, and coarser groups store less but recompute more at once.
Checkpointing trades peak memory for one extra forward pass per block, so it lengthens the backward pass by a fixed amount.

\subsection{The iOS Execution Environment}
\label{sec:background:ios}
Our measurements use MLX~\cite{mlx2023}, Apple's array framework for Apple silicon, which provides the training loop, the quantized weight format and the Metal kernels that execute on the GPU.
The A19 Pro's GPU cores include matrix units, which Apple calls Neural Accelerators~\cite{applea19pro}, and MLX ships kernels that run the quantized matrix multiplication on them.
We call these the accelerated kernels, against the generic kernels that run on the GPU's ordinary arithmetic units.
iOS also constrains the control an application has over its execution.
Four policies impact sustained on-device training.
\paragraph{Memory.}
iOS enforces a per-process memory limit and terminates any process that exceeds it.
On our device the effective limit lies near \SI{6}{\giga\byte} of the \SI{12}{\giga\byte} installed.
Our application holds the increased-memory-limit entitlement~\cite{appleincreasedmemory}, without which iOS terminates training at 750 tokens.
Termination occurs at any point peak memory exceeds the limit, which for training occurs during the first backward pass.
This limit sets the longest sequence a phone can train on.
\paragraph{Frequency and thermal control.}
Dynamic voltage and frequency scaling on iOS is managed entirely by the system.
An application cannot select a frequency or pin work to a core, so it can trade speed for temperature only by idling or through an entitlement that has the system limit its burst performance and keep it at a consistent level~\cite{applesustainedexec}.
The only thermal information exposed is \texttt{ProcessInfo.thermalState}, a four-level indicator without a temperature or a rate of change.
Under sustained load the SoC heats and the system reduces its frequency to hold the package within its thermal envelope.
Inference workloads seldom encounter this regime, since a single request completes within seconds and the device cools between requests.
A training run lasts hours and spends almost all of them throttled, so the throttling behavior determines sustained throughput.
\paragraph{Background execution.}
iOS grants GPU access to background work only through \texttt{BGContinuedProcessingTask} with a GPU resource entitlement~\cite{applebggpu}, and documents that not all devices support this resource.
On our device the scheduler reports no GPU resource, and an ordinary background processing task loses GPU access within seconds of the application leaving the foreground.
Training must therefore run as a foreground application with the display active, so the display draws power throughout every run.
\paragraph{Energy accounting.}
iOS exposes battery state to applications only as \texttt{UIDevice.batteryLevel}, with no access to current, voltage or the power management unit, and our device reported it in steps of five percent of a charge, apart from a final 1\%.
This is the only energy signal an application can read at run time.

%% file: sections/03-configuration.tex
\section{Experimental Setup}
\label{sec:setup}
The previous section described the per-user training stage and the mobile environment training executes in.
This section describes our configuration for our measurements of mobile training performance.
All measurements are taken on an iPhone 17 Pro (model A3523, rated battery energy \SI{55.6}{\kilo\joule}~\cite{eprel17pro}) running MLX~\cite{mlx2023} via \texttt{mlx-swift}, its Swift interface, inside a normal iOS application.
We measure on iOS for two reasons.
First, the iPhone~17~Pro's A19 Pro is among the first phone SoCs with matrix units in the GPU cores~\cite{applea19pro}.
Second, MLX offers a mature array framework for it, with quantized kernels, automatic differentiation and lazy evaluation in one code base, so the training loop runs in the same runtime as inference.
The cost is that iOS restricts applications more than Android, whose foreground services keep work running in the background~\cite{androidforeground}.

\paragraph{Workload:}
The workload we measure is the per-user training stage on the \lamp{} dataset, with our specific model configuration listed in \autoref{tab:config}.
We train on histories from the rating-prediction task of \lamp{}, in which the model predicts the 1-to-5 rating of a user's product review.
They belong to five of the 100 users with the longest histories among those the task adapter was not trained on, and hold 405, 448, 500, 550 and 987 records, spanning the range of those 100 users.
The base model we use is \smollm{}~\cite{smollm3}, chosen because it trains within the memory constraints at the sequence lengths our histories need and performs competitively on standard model benchmarks.
\smollm{} is stored at 4-bit precision with a group size of 64.
We attach adapters of rank 8 to the query and value projections of every block.
A run trains only these adapters, on weights into which the task adapter was merged before quantization.
A training example consists of a history record transformed into the format of the task.
The prompt carries up to four of the user's earlier records, retrieved using BM25~\cite{robertson2009bm25}, a lexical ranking function that scores a record by the query words it shares.
An example of a query is shown in \autoref{fig:example}.
A training run makes three passes over the user's history at a batch size of 1, meaning each training step processes a single example, since activation memory grows with the number of examples in a step.
Examples are truncated at 1024 tokens, the word pieces a model reads text in, the largest length fitting in memory and covering 92\% of the records in these histories.
\autoref{sec:on-device-adapters} evaluates adapter quality on shorter movie-tagging histories instead, with a median of 36 records, which train in a median of 3 minutes.

\paragraph{Measurement:}
One training step consists of preparing the batch, constructing the graph, the forward pass, the backward pass, the optimizer step, and readback.
We time each phase by requesting a result at its boundary, forcing an evaluation.
Because evaluations prevent the runtime from overlapping work, we also time each step without them.
Across the six sequence lengths and both kernels the six phases sum to 1.02 to 1.08$\times$ the fused step.
Throughput is training tokens processed per second.
Where every step has the same length, as in \autoref{sec:memory}, \autoref{sec:time}, \autoref{tab:rank} and the probes of \autoref{sec:sustained:sustained}, steps per second and tokens per second differ by a constant factor, and we report ratios or step times.
Kernel-level breakdowns come from GPU captures around each phase, analyzed later in Xcode's Metal debugger.
The debugger bounds the resources a capture can replay, so a capture of the backward pass samples 12 of the 36 blocks.
Peak memory is read from the runtime allocator at the end of each measurement window.
Energy usage is derived from the battery gauge, which reports charge in 5\% increments.
We report energy net of an idle baseline, the \SI{0.26}{\watt} the phone draws with the screen on and no training.
Unplugged runs start at 85\% charge, except the 987-record run and one 550-record run, which start from a full charge.
Runs whose thermal history could influence one another are separated by at least one hour, longer than the thermal recovery time we measured.
The device ran without a case, screen on at a fixed brightness, with Wi-Fi on and cellular off.

\begin{table}[t]
    \centering\footnotesize
    \caption{On-device training configuration.}
    \label{tab:config}
    \begin{tabular}{@{}ll@{}}
        \toprule
        Model & \smollm{}~\cite{smollm3} \\
        Precision & 4-bit affine, group size 64 \\
        Adapter & LoRA, rank 8, query and value \\
        Adapted blocks & 36 \\
        Optimizer & AdamW~\cite{loshchilov2019adamw}, lr $10^{-5}$, weight decay 0.01 \\
        Batch size & 1 \\
        Gradient checkpointing & per block \\
        Maximum sequence length & 1024 tokens \\
        Steps & \begin{tabular}[t]{@{}l@{}}3 passes over the history,\\ i.e.\ $3 \times$ its record count\end{tabular} \\
        \bottomrule
    \end{tabular}
\end{table}

\begin{figure}[t]
\footnotesize
\begin{verbatim}
[system]
The following are examples of this user's past
activity. Use them to match this user's
preferences and writing style.
- Review: "What a great book with all the
  twists and turn ..." -- the user rated it 5/5
- Review: "..." -- the user rated it 4/5
  (up to 4 retrieved records)
[user]
What is the score of the following review on a
scale of 1 to 5? just answer with 1, 2, 3, 4,
or 5 without further explanation.
review: This book was fantastic!! Such a
touching story of growing up ...
[assistant]
5
\end{verbatim}
\caption{One training example from the rating-prediction task, as the model sees it (shortened). For this user the median example is about 530 tokens and the longest 1190.}
\label{fig:example}
\end{figure}

%% file: sections/04-memory.tex
\section{Memory Analysis of Fine-Tuning}
\label{sec:memory}
Training a 3B-parameter model on mobile devices is possible only because adapters, quantization, and gradient checkpointing bring the memory requirements of fine-tuning within the device's constraints.
This section measures how, and in which combination, to use the aforementioned techniques.

\autoref{tab:memory} reports peak memory usage by sequence length.
Without gradient checkpointing, training \smollm{} peaks at \SI{2.3}{\giga\byte} at 32 tokens and \SI{4.7}{\giga\byte} at 256, and iOS terminates the process on the first backward pass at 512 tokens, a length that 89\% of the records in our histories exceed.
Per-block checkpointing lowers peak memory by 22\% at 32 tokens and 47\% at 256, and reaches 1024 tokens at \SI{4.1}{\giga\byte}.

iOS does not specify the memory limits for a process, but the same sweep over Qwen3-8B~\cite{qwen3} locates the limit.
With per-block checkpointing it trains at sequence lengths from 32 to 256 tokens, peaking at \SI{5.6}{\giga\byte}, before iOS terminates it at 512.
Without checkpointing it fits within the memory constraints at only 32 tokens, using \SI{5.7}{\giga\byte}, which is the largest usage that survived in any configuration and places the limit between it and the smallest terminated run, near \SI{6}{\giga\byte}.
Since only 0.3\% of the records in our histories fit in 256 tokens, Qwen3-8B cannot train on the histories users hold.

\begin{figure}[t]
    \centering
    \includegraphics[width=\linewidth]{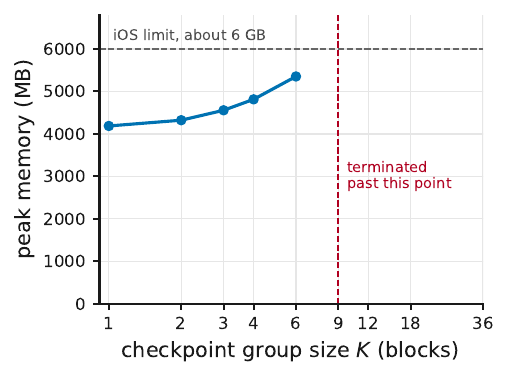}
    \caption{Peak memory against checkpoint granularity in blocks per checkpoint, 1024 tokens. iOS terminated every grouping of nine or more blocks.}
    \label{fig:gc}
\end{figure}

Gradient checkpointing granularity is another lever to control memory usage.
\autoref{fig:gc} plots memory usage against checkpoint granularity and shows that per-block checkpointing yields the lowest memory footprint.
With $L$ blocks checkpointed in groups of $K$, the forward pass retains $L/K$ group boundaries and the backward pass materializes one group's internal activations, so peak memory is $M(K)=aL/K+bK+c$.
A least-squares fit to the five groupings plotted in \autoref{fig:gc} gives $a=\SI{8.1}{\mega\byte}$ per retained boundary, $b=\SI{282}{\mega\byte}$ of internal activations per block and $c=\SI{3.61}{\giga\byte}$ for everything else, at 1024 tokens, and reproduces each measurement within \SI{3}{\mega\byte}.
The minimum lies at $K=\sqrt{aL/b}\approx1$.
A block's internal activation size exceeds its group boundary size by roughly the depth of the model, so the $\sqrt{L}$ grouping that \citet{chen2016gradientcheckpointing} derive for layers of equal memory cost does not apply, and iOS terminates every grouping coarser than six blocks.
None of the coarser groupings that fit trains faster than per-block checkpointing.

Checkpointing recomputes each block's forward pass exactly once during the backward pass, and at matched sequence length it lowers throughput to 0.77--0.85$\times$ that of training without checkpointing.

\begin{table}[t]
    \centering\footnotesize\setlength{\tabcolsep}{3pt}
    \caption{Peak memory (MB) by sequence length. ``Killed'' marks the first length at which iOS terminated the process.}
    \label{tab:memory}
    \begin{tabular}{@{}llrrrrrr@{}}
        \toprule
        \textbf{Model} & \textbf{Checkpointing} & \textbf{32} & \textbf{64} & \textbf{128} & \textbf{256} & \textbf{512} & \textbf{1024} \\
        \midrule
        \smollm{} & none      & 2328 & 2651 & 3312 & 4655 & killed & --- \\
        \smollm{} & per block & 1827 & 1919 & 2109 & 2449 & 3179 & 4125 \\
        Qwen3-8B  & none      & 5671 & killed & --- & --- & --- & --- \\
        Qwen3-8B  & per block & 4712 & 4834 & 5052 & 5560 & killed & --- \\
        \bottomrule
    \end{tabular}
\end{table}

%% file: sections/05-time_breakdown.tex
\section{Time Breakdown and the Backward Path}
\label{sec:time}
This section breaks down a single training step and finds that its backward pass takes twice as long as its arithmetic predicts.
The cause is a defect in a kernel that nothing dispatched, and an audit shows that the kernel the backward pass needs is missing from nine of the ten other runtimes.

\subsection{Per-Kernel Breakdown}
\label{sec:time:breakdown}
A training step consists of six phases, as described in \autoref{sec:setup}.
Before the repair of \autoref{sec:time:repair}, at 250 tokens and above, the backward pass accounts for 78\% of a step, the forward pass 19--22\%, and batch preparation, graph construction, optimizer updates, and readback together account for 1--3\% of a step.
The optimizer step is the only phase that deals exclusively with adapter state, and it makes up less than 2\% of training time.

Within the two passes, most of the time is spent within the frozen base model, as shown by the per-kernel breakdown in \autoref{fig:perop_breakdown}.
Quantized matrix multiplication over the frozen weights is 86\% of the forward pass and 90\% of the backward pass.
The adapter's low-rank multiplications are 3\% and 2\%, respectively.
Attention, normalization and rotary embedding together are under 1\% of either pass.
Thus, the time spent training at these sequence lengths is driven by the weight matrices, not attention.
Adding together the adapter's forward and backward products with the optimizer update gives \SI{0.15}{\second} out of a \SI{3.55}{\second} step at 250 tokens, roughly 4\%.
The remaining 96\% of the time is spent on the frozen model, and quantized matrix multiplication alone takes 87\% of the step.
Parameter-efficient fine-tuning on mobile devices therefore reduces the memory of a training step, but concentrates the work within the frozen model.

We note that the backward pass is more expensive than it should be.
The backward pass computes one input-gradient per block plus one forward pass recomputation due to checkpointing, so arithmetic alone predicts the backward pass to take twice as long as the forward pass.
At 250 tokens and above it takes 3.5--4.2$\times$ as long, and up to 4.4$\times$ at 100 tokens.
Disabling checkpointing at 100 to 375 tokens, where training fits without it, separates the two terms. Recomputation costs the time of 0.7 to 0.9 forward passes, and the input-gradient products cost 3.4 to 3.6.
The excess time is concentrated in the quantized matrix multiplication of the backward pass.

\begin{widefigure}[t]
    \centering
    \includegraphics[width=\linewidth]{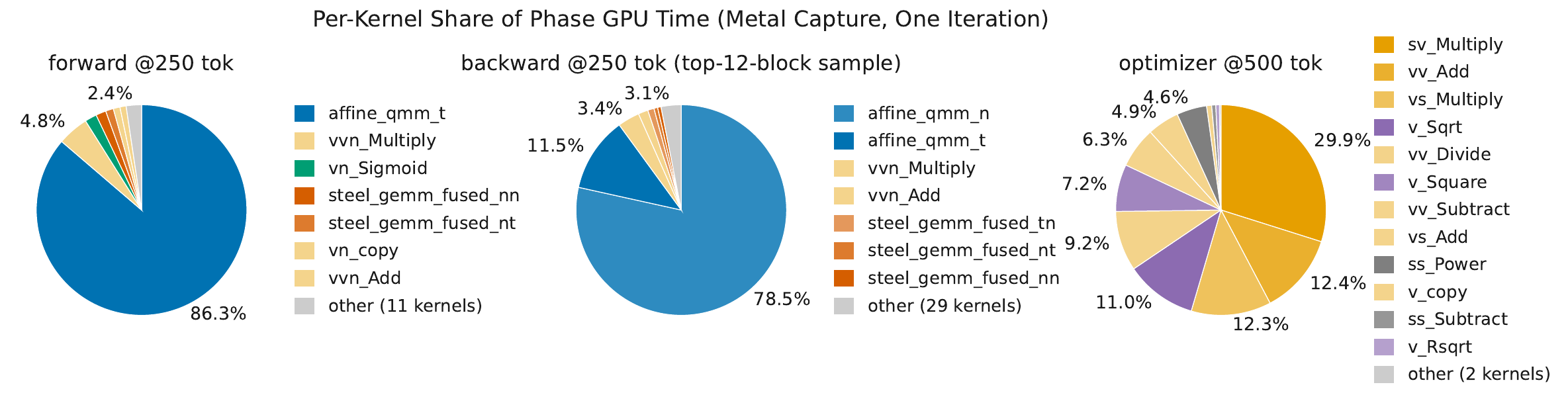}
    \caption{Per-kernel share of GPU time within each phase of a training step before the kernel repair, at 250 tokens and at 500 for the optimizer update. \texttt{affine\_qmm\_t} and \texttt{affine\_qmm\_n} are the quantized products of the forward and backward directions, and \texttt{affine\_qmm\_t} also appears in the backward pass as checkpoint recomputation.}
    \label{fig:perop_breakdown}
\end{widefigure}

\subsection{Backward Pass Diagnosis}
\label{sec:time:backward_pass}
The forward and backward pass multiply by the same frozen weight matrices, so the excess time lies in how the backward pass reads them.
A quantized weight stores a 4-bit code together with a scale and offset shared by a group of $G=64$ consecutive weights along the input dimension of the layer.
For a linear layer with input $X\in\mathbb{R}^{M\times N}$ and weights $W\in\mathbb{R}^{N\times K}$, the forward pass $Y=XW$ reads $N$ weights down one column of $W$, which is $N/G$ whole groups.
Writing $g$ for the group containing index $n$ and $q$ for the stored codes,
\begin{equation*}
    Y_{m,k} = \sum_{g} \Big( s_{g,k} \sum_{n \in g} X_{m,n}\, q_{n,k} + \beta_{g,k} \sum_{n \in g} X_{m,n} \Big).
\end{equation*}
The inner sums run over integer codes, and the scale is applied once per group of 64 products.

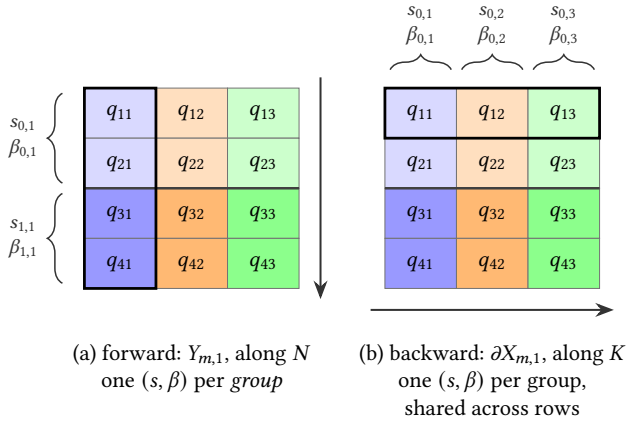
\begin{figure}[t]
    \centering
    \resizebox{\figsmall}{!}{\input{figures/qmm}}
    \caption{Group layout and the two contraction directions over the same stored $W$. Fill color identifies the group, and the outline marks the slice one output element reads. Drawn with $N{=}4$, $K{=}3$, $G{=}2$, whereas $G{=}64$ in our configuration.}
    \label{fig:qmm}
\end{figure}

The backward pass computes the input gradient $\partial X=\partial Y\,W^{\top}$ and reads $W$ the other way, as \autoref{fig:qmm} shows.
One element $\partial X_{m,n}=\sum_k \partial Y_{m,k}\,w_{n,k}$ reads one row of $W$, whose weights all lie in one group but carry a different scale $s_{g,k}$ in every column, so the scale no longer factors out of the sum over $k$.
It factors out along $n$ instead, since the 64 rows of a group share their scales, so a kernel for this contraction scales a tile of $\partial Y$ once per group and reuses it across the group's rows.
The arithmetic is the same as in the forward pass, but the codes are packed along $N$, so this kernel reads them across packed words and needs a tiling of its own.

The backward pass therefore needs a second kernel, and MLX has one.
An accelerated kernel for this contraction exists in the source and is compiled into every binary, but no code path dispatches it.
The backward pass instead runs a generic Metal kernel on the GPU's ordinary arithmetic units, which makes it 1.65 to 2.13 times slower than with the accelerated kernel (\autoref{sec:time:repair}).

\subsection{Repairing the Backward Kernel}
\label{sec:time:repair}
We found two defects in the accelerated kernel at our group size of 64. It indexed the stored weights and scales with offsets computed for the transposed layout, and it did not bound its tile loads, so it read and wrote out of bounds whenever a matrix's row count was not a multiple of the tile height.
The repair corrects the offsets and leading dimension passed to the tile loader, bounds the loads as the forward kernel does, and dispatches the kernel in place of the generic one.\anon[]{\footnote{The repair was merged into MLX as pull request \#4051 (\url{https://github.com/ml-explore/mlx/pull/4051}), and MLX dispatches the repaired kernel by default since release 0.32.3.}}
The kernel entered MLX together with its support for the matrix units and was unchanged until this work.
MLX's quantized linear layer computes only the transposed product, so the forward direction was wired to the matrix units, while the backward direction, which training needs at every step, was not.
Every defect reported against these kernels before ours concerned an inference path.
We first alternate the generic and repaired kernels between successive steps at six sequence lengths from 50 to 1000 tokens, so that both see the same thermal state.
The backward pass runs 1.65 to 2.13$\times$ faster and the whole step 1.45 to 1.68$\times$ faster, 1.55$\times$ on average per step.
Trained unplugged from 85\% charge with the same batch order under each kernel, a complete adapter over the 405-record history takes \SI{2.55}{\hour} under the generic kernel and \SI{1.74}{\hour} under the repaired one, a factor of 1.47, and draws \SI{39.3}{\kilo\joule} instead of \SI{26.2}{\kilo\joule} net of the idle baseline.
Across the six lengths the backward pass now takes 1.8 to 2.2 times the forward pass, against 3.4 to 4.4 before the repair, so the generic kernel accounts for the whole excess over the factor of two that its arithmetic predicts.
The repaired kernel agrees bit for bit with a dense matrix multiplication over the dequantized weights on the shapes where that reference runs on the matrix units, and to within $10^{-3}$ relative error elsewhere.
The two runs' loss trajectories coincide, with per-report losses correlated at 0.9999 and mean loss over the final 50 steps of 0.7911 under the generic kernel and 0.7820 under the repaired one.
The adapter's share of the training step rises from 4\% to 7\%, with the frozen base still dominating.

\subsection{ML Runtime Audit}
\label{sec:time:audit}
Every runtime that stores quantized weights needs this second kernel to compute the backward contraction without dequantizing the weights.
We examined the source code of eleven inference and training runtimes, as it stood between 17 and 20 August 2026, and determined, for each, whether an accelerated quantized matrix multiplication exists for that contraction and whether a training loop dispatches it.
We count a kernel as accelerated if it is a GPU kernel tiled for this contraction over the packed low-bit weights that, on a GPU with matrix units, runs on them, which excludes generic kernels such as the one MLX dispatched before the repair.

\autoref{tab:audit} shows the results.
Six runtimes have no quantized backward pass at all, and where they support training they do so on floating-point weights.
Three compute the input gradient by dequantizing the weights for a dense multiplication, ONNX Runtime and bitsandbytes~\cite{bitsandbytesrepo} over the whole matrix and llama.cpp row by row on the CPU.
On our device, dequantizing each weight matrix for a dense multiplication on the matrix units makes the backward pass 12 to 44\% slower than the repaired kernel at 250 to 1024 tokens and raises peak memory by 1.1 to \SI{1.5}{\giga\byte}.
One runtime, ExecuTorch~\cite{executorch2026}, added an accelerated backward kernel for quantized weights to its Vulkan and WebGPU backends in July 2026~\cite{executorchq4gsw}, concurrently with this work.
MLX was the only other runtime in which such a kernel existed, and it was never dispatched and incorrect until the repair.
For inference, all eleven store quantized weights in the layout the forward pass reads, and all eleven have optimized that read.
The contraction that runs the other way is needed only by training, which on-device runtimes built for inference have had little reason to accelerate.

\begin{table}[t]
  \centering\footnotesize\setlength{\tabcolsep}{3pt}
  \caption{Input-gradient path over quantized weights in eleven runtimes.}
  \label{tab:audit}
  \begin{tabular}{@{}lll@{}}
    \toprule
    \textbf{Runtime} & \textbf{Accelerated kernel} & \textbf{Training path} \\
    \midrule
    ExecuTorch~\cite{executorch2026} & yes & accelerated kernel \\
    MLX~\cite{mlx2023} & yes, not dispatched & generic kernel \\
    \addlinespace
    ONNX Runtime~\cite{onnxruntime} & no & dequantize, dense \\
    bitsandbytes~\cite{bitsandbytesrepo} & no & dequantize, dense \\
    llama.cpp~\cite{llamacpp2023} & no & dequantize per row, CPU \\
    \addlinespace
    MNN~\cite{mnn2020} & no & float weights only \\
    LiteRT~\cite{litert} & no & float weights only \\
    TVM~\cite{tvm2018} & no & float weights only \\
    Core ML~\cite{coreml} & no & float weights only \\
    MLC-LLM~\cite{mlcllm} & no & no training \\
    ncnn~\cite{ncnn} & no & no training \\
    \bottomrule
  \end{tabular}
\end{table}

%% file: figures/qmm.tex
\begin{tikzpicture}[
    font=\small,
    cell/.style ={draw=black!55, minimum width=10mm, minimum height=7mm,
                  inner sep=0pt},
    g0a/.style  ={cell, fill=blue!15},   g1a/.style={cell, fill=blue!40},
    g0b/.style  ={cell, fill=orange!25}, g1b/.style={cell, fill=orange!55},
    g0c/.style  ={cell, fill=green!20},  g1c/.style={cell, fill=green!45},
    gline/.style={draw=black!75, line width=1pt},
    sel/.style  ={draw=black, line width=1.1pt},
    ax/.style   ={-{Stealth[length=2.6mm]}, thick, black!80},
    br/.style   ={decorate, decoration={brace, amplitude=6pt}, semithick,
                  black!80},
    head/.style ={font=\small, align=center, anchor=north},
    take/.style ={font=\small, align=center, anchor=north},
    lab/.style  ={font=\footnotesize, align=center},
]

\newcommand{\wgrid}[1]{%
  \node[g0a] (#1-11) at (10mm,  -7mm) {$q_{11}$};
  \node[g0b] (#1-12) at (20mm,  -7mm) {$q_{12}$};
  \node[g0c] (#1-13) at (30mm,  -7mm) {$q_{13}$};
  \node[g0a] (#1-21) at (10mm, -14mm) {$q_{21}$};
  \node[g0b] (#1-22) at (20mm, -14mm) {$q_{22}$};
  \node[g0c] (#1-23) at (30mm, -14mm) {$q_{23}$};
  \node[g1a] (#1-31) at (10mm, -21mm) {$q_{31}$};
  \node[g1b] (#1-32) at (20mm, -21mm) {$q_{32}$};
  \node[g1c] (#1-33) at (30mm, -21mm) {$q_{33}$};
  \node[g1a] (#1-41) at (10mm, -28mm) {$q_{41}$};
  \node[g1b] (#1-42) at (20mm, -28mm) {$q_{42}$};
  \node[g1c] (#1-43) at (30mm, -28mm) {$q_{43}$};
  \draw[gline] (5mm,-17.5mm) -- (35mm,-17.5mm);
}

\begin{scope}[local bounding box=A]
  \wgrid{a}
  \draw[sel] (5mm,-3.5mm) rectangle (15mm,-31.5mm);      %
  \draw[br] ($(a-21.south west)+(-3mm,0.8mm)$) -- ($(a-11.north west)+(-3mm,-0.8mm)$)
    node[midway, left=6pt, lab] {$s_{0,1}$\\$\beta_{0,1}$};
  \draw[br] ($(a-41.south west)+(-3mm,0.8mm)$) -- ($(a-31.north west)+(-3mm,-0.8mm)$)
    node[midway, left=6pt, lab] {$s_{1,1}$\\$\beta_{1,1}$};
  \draw[ax] (38mm, -2mm) -- (38mm, -33mm);
  \node[take] at (20mm, -38mm)
        {(a) forward: $Y_{m,1}$, along $N$\\one $(s,\beta)$ per \emph{group}};
\end{scope}

\begin{scope}[xshift=42mm, local bounding box=B]
  \wgrid{b}
  \draw[sel] (5mm,-3.5mm) rectangle (35mm,-10.5mm);      %
  \foreach \k in {1,2,3} {
    \draw[br] ($(b-1\k.north west)+(0.8mm,2.5mm)$) -- ($(b-1\k.north east)+(-0.8mm,2.5mm)$)
      node[midway, above=6pt, lab] {$s_{0,\k}$\\$\beta_{0,\k}$};
  }
  \draw[ax] (3mm, -34.5mm) -- (37mm, -34.5mm);
  \node[take] at (20mm, -38mm)
        {(b) backward: $\partial X_{m,1}$, along $K$\\one $(s,\beta)$ per group,\\shared across rows};
\end{scope}

\end{tikzpicture}

%% file: sections/07-systems_measure.tex
\section{End-to-End Adapter Training}
\label{sec:sustained}
Prior work has measured single training steps of multi-billion-parameter models on phones, but a per-user adapter takes thousands of steps, over which the phone throttles and drains its battery.
We therefore train complete adapters on the repaired runtime over user histories of 405 to 987 records, and measure their training time (\autoref{sec:sustained:time}), the cost of the adapter configuration (\autoref{sec:cost:config}), throughput under sustained load (\autoref{sec:sustained:sustained}), and energy (\autoref{sec:cost:energy}).

\subsection{Time}
\label{sec:sustained:time}
Because attention takes under 1\% of a step at these lengths (\autoref{sec:time:breakdown}), the cost of a run should depend on the number of training tokens it processes rather than on the number of records.
\autoref{tab:cost} lists the five end-to-end runs ordered by token counts.
Wall-clock time is linear in tokens.
We fit training time as $t = 0.0112\,\mathrm{s}\cdot\text{tokens} + 13\,\mathrm{s}$, which has $r=0.998$ and no residual above 6\%, against $r=0.931$ and a residual of 35\% for a fit on record count.
The 448-record history takes longer than the 500- and 550-record histories because its records are longer.
A run's duration can therefore be estimated from its token count before it starts, and the per-token rate of 0.011 to 0.012 seconds is the same across histories.
Repeating the 405-record run three times gave 6872, 6948 and 6616 seconds, a spread of 4.9\%.
The three runs agree within 1\% on their cold-start rate, the rate over their first ten steps, and differ in the level their throughput settles to under sustained load.
The same history also trains faster unplugged. One run on battery power took \SI{1.74}{\hour}, 5 to 10\% less than the plugged runs, with a slightly slower cold start and a faster plateau, so the difference arises under sustained load.

\begin{table}[t]
    \centering
    \caption{Complete adapter training runs, plugged in, ordered by token count. Tokens count all passes over the history.}
    \label{tab:cost}
    \begin{tabular}{S[table-format=3.0] S[table-format=1.2] S[table-format=5.0] S[table-format=1.4]}
        \toprule
        {\textbf{Records}} & {\textbf{Tokens (M)}} & {\textbf{Time (s)}} & {\textbf{s/token}} \\
        \midrule
        405 & 0.64 &  6812 & 0.0106 \\
        550 & 1.11 & 12443 & 0.0112 \\
        500 & 1.16 & 13080 & 0.0113 \\
        448 & 1.27 & 14924 & 0.0118 \\
        987 & 2.35 & 26099 & 0.0111 \\
        \bottomrule
    \end{tabular}
\end{table}

\subsection{Adapter Configuration}
\label{sec:cost:config}
Our recipe fixes three properties of the adapter: its rank, the projections it attaches to, and the blocks it covers.
Each is a knob to trade model quality against cost, and we seek to find which of them cost depends on.
\autoref{tab:rank} varies each on a device heat-soaked by 45 minutes of continuous training and reports throughput, FLOP count, and memory usage relative to the original recipe (first row), with the depth rows compared against the recipe measured in the same session.
Each cell is the mean of three measurements.

Raising rank from 8 to 32 on the same two projections costs 8 to 10\% of throughput while quadrupling the adapter's parameters. The low-rank products add little arithmetic, and the number of kernels launched stays fixed.
Changing which projections have adapters matters more, as each adds an extra kernel launch.
Adapting all seven linear projections of a block instead of only the query and value projections cost 17 to 27\% of throughput, and rank 64 on all seven projections at 512 tokens degrades throughput by 35\% with \SI{1.3}{\giga\byte} of additional memory used.
Even then, the adapter's forward products are 4.3\% of the frozen model's arithmetic at 512 tokens.
Rather, adapters of rank 8 or 64 fail to fill the GPU's matrix units, resulting in low-rank products running an order of magnitude less efficiently per operation than the quantized base matrix multiplication.

Adapter depth changes the amount of computation done, since backpropagation stops at the first block that holds a trainable parameter, as in layer freezing~\cite{ruckle2021adapterdrop,wang2023egeria}.
Every block before the first trainable parameter runs its forward pass and nothing else.
Adapting the last 18 of 36 blocks takes 0.60$\times$ the time of adapting all 36 at 512 tokens and 0.57$\times$ at 1024.
Peak memory falls by at most 6\%, since the frozen weights and the checkpointed activations of every block stay resident whether or not the block is adapted.

Rank, projections, and depth also determine what the adapter can learn.
We measured this for depth on a server, at 4-bit precision, with the per-user movie-tagging adapters of \autoref{sec:quality:per-user}, over the 100 users and 3,302 queries evaluated there.
Adapters on the last 24 blocks reach an accuracy of 0.476, against 0.488 on all 36, retaining 91\% of the gain over the task adapter across queries and 66\% of the mean gain per user.
Adapters on the last 18 blocks reach 0.472, retaining 88\% and 64\%.
Adapting the last 24 blocks raises throughput on the phone by 34 to 45\% in \autoref{tab:rank} at a cost of 1.1 accuracy points.
The rest of the paper keeps the adapter shape of the per-user recipe of \citet{oppu2024}, so that its measurements describe the published adapter.

\begin{table}[t]
    \centering\footnotesize\setlength{\tabcolsep}{3pt}
    \caption{Cost of the adapter configuration at the throttled plateau.}
    \label{tab:rank}
    \begin{tabular}{@{}rlrrrrr@{}}
        \toprule
        & & & \textbf{Adapter} & \multicolumn{2}{c}{\textbf{Throughput vs.\ recipe}} & \textbf{Peak} \\
        \cmidrule(lr){5-6}
        \textbf{Rank} & \textbf{Proj.} & \textbf{Blocks} & \textbf{FLOPs} & \textbf{512 tok} & \textbf{1024 tok} & \textbf{(MB)} \\
        \midrule
        8  & q, v   & 36 & 0.1\% & --- & --- & 4185 \\
        32 & q, v   & 36 & 0.3\% & $-8\%$ & $-10\%$ & 4252 \\
        8  & all 7  & 36 & 0.5\% & $-26\%$ & $-17\%$ & 4341 \\
        32 & all 7  & 36 & 2.1\% & $-27\%$ & $-18\%$ & 4859 \\
        64 & all 7  & 36 & 4.3\% & $-35\%$ & $-19\%$ & 5552 \\
        \addlinespace
        8  & q, v   & 30 & 0.1\% & $+15\%$ & $+18\%$ & 4009 \\
        8  & q, v   & 24 & 0.1\% & $+34\%$ & $+45\%$ & 4077 \\
        8  & q, v   & 18 & 0.1\% & $+67\%$ & $+77\%$ & 4040 \\
        8  & q, v   & 12 & 0.1\% & $+106\%$ & $+134\%$ & 4020 \\
        8  & q, v   &  6 & 0.1\% & $+171\%$ & $+243\%$ & 3927 \\
        \bottomrule
    \end{tabular}
\end{table}

\subsection{Sustained Training}
\label{sec:sustained:sustained}
Every measured training run spends most of its time in a throttled thermal state.
Every one starts at a cold-start rate and loses about half of it, most of that within minutes, which is why a per-token rate measured in a run's first minutes overstates what the run will achieve by about a factor of two.

\begin{figure}[t]
    \centering
    \includegraphics[width=\linewidth]{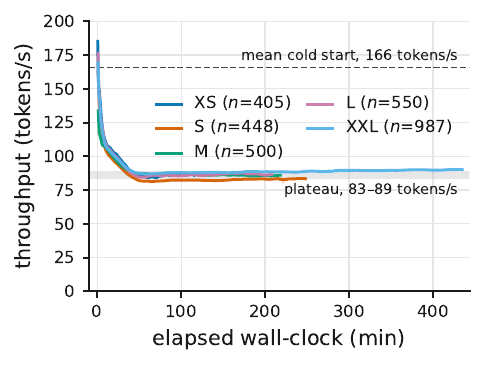}
    \caption{Throughput of the five complete adapter runs of \autoref{tab:cost} against wall-clock time. Faint traces are per-window values over ten steps, bold traces a centered rolling mean.}
    \label{fig:thermal_collapse}
\end{figure}

\begin{figure}[t]
    \centering
    \includegraphics[width=\linewidth]{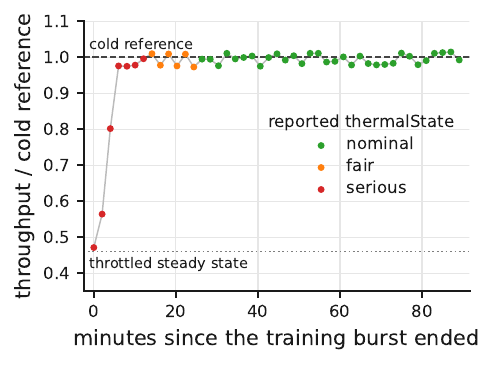}
    \caption{Throughput recovery after an hour of continuous training, probed with a 500-token step every two minutes and normalized to the cold rate. Color is the value of \texttt{ProcessInfo.thermalState} at each probe.}
    \label{fig:thermal_recovery}
\end{figure}

\autoref{fig:thermal_collapse} shows the throughput of the five complete runs in training tokens per second.
Whatever history it trains on, every plugged-in run settles to between 83 and 89 tokens per second, about half of the 166 tokens per second of a cold start, and none recovers for the rest of the run.
A plateau that does not depend on the history is what makes the time of a run linear in its tokens (\autoref{sec:sustained:time}).
Half of the throughput loss occurs within about six minutes, and throughput comes within 5\% of its plateau after 32 to 38 minutes.
The pattern is consistent with the SoC reaching its temperature limit and iOS lowering the clock frequency to hold it there, and the plateau is the rate the device sustains at that limit.
Throttling slows all phases of a step alike. After a 45-minute soak, a measurement of the 500-token step that forces evaluation at every phase boundary (\autoref{sec:setup}), and so idles the GPU between phases, finds it 1.41$\times$ slower than its cold rate, and the share of each of its six phases moves by at most 1.5 percentage points, so the breakdown of \autoref{sec:time} holds throughout a run.
Continuous training, which does not idle, is throttled more deeply, to 1.9 to 2.0$\times$ its cold-start step time in the runs of \autoref{fig:thermal_collapse} and to 2.17$\times$ in the session of \autoref{fig:thermal_recovery}.

Once training stops, single-step throughput recovers within minutes, but iOS's thermal signal does not track it.
\autoref{fig:thermal_recovery} probes throughput with a single step every two minutes after an hour of continuous training at 500 tokens that had slowed the step by 2.17$\times$ from its cold rate.
Between probes the device is idle, so this is the recovery of an idle device.
Throughput regains half of its loss within four minutes and 95\% within six, the resolution of probes two minutes apart.
The indicator \texttt{ProcessInfo.thermalState}, the only thermal information iOS exposes, does not return to \texttt{nominal} until 26 minutes have passed, more than four times longer than throughput takes to recover.

Since throughput recovers within minutes, an application could try to train in bursts or to pace its steps and spend part of the run cool.
We tested both on a heat-soaked device in thirteen interleaved twenty-minute blocks, alternating continuous training with a one-second pause after every step, a three-second pause, and ten minutes of training followed by two minutes of rest.
Every schedule completed fewer steps per hour than continuous training, ranging from 0.91 to 0.92$\times$ with the one-second pause, 0.78 to 0.79$\times$ with the three-second pause, to 0.88 to 0.89$\times$ in bursts.
Pausing does cool the device, as the three-second pause shortens the active step from 5.2 to \SI{3.6}{\second}, but each second of pause buys back about half a second of compute, and a burst that starts after two minutes of rest returns to the continuous step time within its first minutes.
Training already runs at the power the chassis can dissipate. If power grows faster than linearly with frequency, as it does under voltage scaling, a burst at a higher frequency does less work per joule than continuous training at the throttled one, so time spent idle is not recovered.

\subsection{Energy}
\label{sec:cost:energy}
We repeated three of the histories unplugged to measure the energy of a complete adapter.

\autoref{tab:energy} reports the four runs.
The 405-record history, started at 85\%, completed after drawing \SI{26.2}{\kilo\joule} net of the idle baseline, 47\% of a charge.
The 550-record history, also started at 85\%, shut the device down at step 1590 of 1650 after drawing \SI{43.9}{\kilo\joule}, 79\% of a charge, and a second attempt shut down at step 1630.
Started from a full charge, the same history completed after drawing \SI{44.4}{\kilo\joule}, 80\% of a charge.
The 987-record history, started from a full charge, shut the device down at step 1640 of 2961 after drawing \SI{51.6}{\kilo\joule}, 93\% of the rated capacity.
A full charge therefore gives training about \SI{52}{\kilo\joule}, and the device shuts down at a reported 1\%.

Net power during training is 4.0 to \SI{4.2}{\watt} across runs that differ in user history, duration and starting charge, and the generic kernel of \autoref{sec:time:repair} draws a similar \SI{4.3}{\watt}, so the repair saves energy in proportion to the time it saves.
With power constant, energy per token is constant too, 0.04 joules in every run, and the token count that predicts a run's time in \autoref{sec:sustained:time} predicts its energy as well.
A full charge trains about 1.3 million training tokens.
An adapter trained with our recipe thus costs about 0.01 seconds and 0.04 joules per training token, and three times that per token of history, over which the recipe makes three passes.

\begin{table}[t]
    \centering\footnotesize\setlength{\tabcolsep}{4pt}
    \caption{Energy to train one adapter unplugged, net of the idle baseline.}
    \label{tab:energy}
    \begin{tabular}{@{}S[table-format=3.0] S[table-format=3.0] S[table-format=1.2] S[table-format=2.1] S[table-format=2.1] S[table-format=1.2] S[table-format=1.1] l@{}}
        \toprule
        & {\textbf{Start}} & & {\textbf{Net}} & {\textbf{\%}} & & {\textbf{Net}} & {\textbf{Steps}} \\
        {\textbf{Records}} & {\textbf{\%}} & {\textbf{Hours}} & {\textbf{kJ}} & {\textbf{charge}} & {\textbf{J/token}} & {\textbf{W}} & {\textbf{done}} \\
        \midrule
        405 &  85 & 1.74 & 26.2 & 47.1 & 0.04 & 4.2 & 1215/1215 \\
        550 &  85 & 3.01 & 43.9 & 78.9 & 0.04 & 4.1 & 1590/1650 \\
        550 & 100 & 3.07 & 44.4 & 79.8 & 0.04 & 4.0 & 1650/1650 \\
        987 & 100 & 3.63 & 51.6 & 92.9 & 0.04 & 4.0 & 1640/2961 \\
        \bottomrule
    \end{tabular}
\end{table}

%% file: sections/08-quality.tex
\section{Personalization Quality}
\label{sec:quality}
This section evaluates the task adapter against prompt retrieval (\autoref{sec:quality:ftvsretrieve}), measures what a per-user adapter adds to the task adapter (\autoref{sec:quality:per-user}), and repeats the per-user experiment with training and inference on the phone (\autoref{sec:on-device-adapters}).

\subsection{Setup}
\label{sec:quality:setup}
We use the seven \lamp{}~\cite{lamp2024} tasks selected by \citet{oppu2024}, four classification tasks (citation identification, movie tagging, news categorization, rating prediction) and three generation tasks (news headline, scholarly title, and tweet paraphrase).
Classification tasks are scored using accuracy, with mean absolute error (MAE) also reported for rating prediction, while generation tasks are scored using \rougeone{}~\cite{lin2004rouge}, which rates from 0 to 1 how many words the generated text shares with the text the user actually wrote.
A task adapter is trained once on many users' records to teach the model a task, and a per-user adapter is then trained on top of it from one user's history (\autoref{sec:ft-personalization}).
The benchmark's original test split contains one or two queries per user for most tasks, enough to measure whether a task adapter helps users in general.
To measure a per-user adapter, we follow \citet{oppu2024} and evaluate on the users with the longest histories per task, who on three of the seven tasks contribute tens to hundreds of queries each.

\subsection{Fine-Tuning vs. Retrieval}
\label{sec:quality:ftvsretrieve}
We compare four ways of serving a user, including whether to fine-tune or not, and whether to use retrieval or not.
The base model with no retrieval is the baseline.
Retrieval places the four most relevant records in the prompt and requires no training.
One task adapter is trained per task on the records of the training split users, so every evaluated user is unseen and served with either no retrieved history or with the four retrieved records.

Results are shown in \autoref{tab:ft-vs-prompt}.
Without any retrieved records, the task adapter outperforms the base model with retrieval on six of the seven tasks, by 3 to 19 points, with prompts two to five times shorter.
The largest gain is on movie tagging, where accuracy rises from 0.278 with retrieval to 0.468 with the adapter.
The exception is citation identification, in which the model must decide which of two references the user cited and the only signal is the user's past papers.
On this task the base model scores 0.500, exactly the same as randomly guessing, and the adapter alone reaches 0.536, below the 0.549 of retrieval.
Fine-tuning and retrieval are complementary, and the adapter with retrieval is best on every task.
Adding retrieval to the adapter raises citation identification from 0.536 to 0.716 and the other six tasks by 2 to 10 points.
Retrieval on top of an adapter therefore trades a longer prompt and its computation for further quality.

\begin{widetable}[t]
    \centering\small\setlength{\tabcolsep}{4pt}
    \caption{Benchmark-wide evaluation on the \lamp{} test split at 4-bit precision, where each test item is one user. Retr.\ places the four most similar records in the prompt, and Adapter is the task adapter. $\Delta$ columns are relative to Base+Retr.}
    \label{tab:ft-vs-prompt}
    \begin{tabular*}{\textwidth}{@{\extracolsep{\fill}}ll r S[table-format=1.3] S[table-format=1.3] S[table-format=1.3] S[table-format=+1.3,retain-explicit-plus] S[table-format=1.3] S[table-format=+1.3,retain-explicit-plus]@{}}
        \toprule
        \textbf{Task} & \textbf{Metric} & {\textbf{\boldmath$n$}} & {\textbf{Base}} & {\textbf{Base+Retr.}} & {\textbf{Adapter}} & {\textbf{\boldmath$\Delta$}} & {\textbf{Adapter+Retr.}} & {\textbf{\boldmath$\Delta$}} \\
        \midrule
        Citation identification & \acc\up      & 2500 & 0.500 & 0.549 & 0.536 & -0.012 & \best{0.716} & +0.167 \\
        Movie tagging           & \acc\up      &  870 & 0.136 & 0.278 & 0.468 & +0.190 & \best{0.547} & +0.269 \\
        News categorization     & \acc\up      & 1274 & 0.256 & 0.542 & 0.646 & +0.104 & \best{0.741} & +0.199 \\
        Rating prediction       & \acc\up      & 2500 & 0.437 & 0.648 & 0.712 & +0.064 & \best{0.784} & +0.136 \\
                                & \mae\down    & 2500 & 0.684 & 0.442 & 0.331 & -0.111 & \best{0.243} & -0.200 \\
        \addlinespace
        News headline           & \rougeone\up & 2376 & 0.136 & 0.148 & 0.176 & +0.028 & \best{0.211} & +0.063 \\
        Scholarly title         & \rougeone\up & 2500 & 0.331 & 0.336 & 0.455 & +0.118 & \best{0.482} & +0.146 \\
        Tweet paraphrase        & \rougeone\up & 1496 & 0.395 & 0.414 & 0.516 & +0.102 & \best{0.535} & +0.120 \\
        \bottomrule
    \end{tabular*}
\end{widetable}

\subsection{Per-User Adaptation}
\label{sec:quality:per-user}
The task adapter in the previous section learns the structure of a task and nothing about the user.
From the user's own records, a per-user adapter is trained on top of it.
Our second question is whether this adapter adds anything.
Measuring this requires many queries from the same user, so we use the long-history users of \citet{oppu2024} and reproduce their pipeline, where each user's most recent records are the test queries and the earlier ones form the history the adapter is trained on.
Their pipeline uses its own task adapters, prompt format and a single retrieved record, so its scores are not comparable with those of \autoref{tab:ft-vs-prompt}.
These comparisons run on a server at 4-bit, the precision the phone trains in.

\autoref{tab:per-user} shows the results.
$\Delta$ is the gain over all queries, which weights each user by the number of queries they contribute, and $\Delta_u$ is the mean of per-user gains, which weights every user equally.
The per-user adapter improves the score on five of the seven tasks, but the gain is large only on movie tagging, where accuracy rises from 0.361 to 0.488, a $\Delta$ of 12.7 points and a $\Delta_u$ of 3.0.
On the other two tasks with tens of queries per user the gain is below one point, with news categorization rising from 0.805 to 0.811 and news headline generation from 0.158 to 0.159.
The four remaining tasks contribute about one query per user, and among them the per-user adapter lowers the MAE of rating prediction from 0.223 to 0.205 and raises citation identification by 1.5 points per user, while tweet paraphrase and scholarly title drop by less than one point.
To check that the gain comes from the user's own history and not just from the extra training, we also trained each user's adapter on another user's history instead of their own.
That adapter adds 1.2 points over all queries and 1.0 per user, against 12.7 and 3.0 for the user's own history, so most of the movie-tagging gain is specific to the user.

\begin{widetable}[t]
    \centering\small\setlength{\tabcolsep}{4pt}
    \caption{Per-user evaluation on the long-history users of \citet{oppu2024} at 4-bit precision, with the same queries in both arms. Task is a task adapter trained with their released code and evaluated with one retrieved record, and +User adds a per-user adapter trained with the recipe of \autoref{tab:config}.}
    \label{tab:per-user}
    \begin{tabular*}{\textwidth}{@{\extracolsep{\fill}}ll r r S[table-format=1.3] S[table-format=1.3] S[table-format=+1.3,retain-explicit-plus] S[table-format=+1.3,retain-explicit-plus]@{}}
        \toprule
        \textbf{Task} & \textbf{Metric} & {\textbf{Users}} & {\textbf{Test q/user}} & {\textbf{Task}} & {\textbf{+User}} & {\textbf{\boldmath$\Delta$}} & {\textbf{\boldmath$\Delta_u$}} \\
        \midrule
        News categorization     & \acc\up      &  49 & 123 & 0.805 & \best{0.811} & +0.006 & +0.012 \\
        News headline           & \rougeone\up & 100 &  67 & 0.158 & \best{0.159} & +0.001 & +0.001 \\
        Movie tagging           & \acc\up      & 100 &  33 & 0.361 & \best{0.488} & +0.127 & +0.030 \\
        Movie tagging, on phone & \acc\up      & 100 &  33 & 0.366 & \best{0.486} & +0.120 & +0.027 \\
        \addlinespace
        Citation identification & \acc\up      & 100 & 1.2 & \best{0.634} & \best{0.634} & 0.000 & +0.015 \\
        Rating prediction       & \mae\down    & 101 & 1.1 & 0.223 & \best{0.205} & -0.018 & -0.025 \\
        Tweet paraphrase        & \rougeone\up & 100 & 1.1 & \best{0.535} & 0.530 & -0.005 & -0.003 \\
        Scholarly title         & \rougeone\up & 101 & 1.1 & \best{0.494} & 0.486 & -0.008 & -0.006 \\
        \bottomrule
    \end{tabular*}
\end{widetable}

\subsection{On-Device Adapters}
\label{sec:on-device-adapters}
The per-user results of \autoref{sec:quality:per-user} were measured on a server, which uses the same 4-bit weights as the phone but different kernels and higher-precision arithmetic.
To validate them, we repeated the movie tagging experiment entirely on the device, with the per-user adapter trained on the phone and all predictions generated there.

\autoref{tab:per-user} lists the result as the ``on phone'' row, directly below the server's movie-tagging row, over the same 100 users and 3,302 test queries.
The phone-trained and the server-trained adapter agree within one accuracy point and differ on 7.8\% of individual predictions.
The per-user gain on the device (+12.0 over all queries, +2.7 per user) is close to that on the server (+12.7, +3.0).
Training the 100 adapters took a median of 3 minutes per user, on histories of 21 to 774 records with a median of 36.
A linear fit of their training times against tokens has a slope within 1\% of the per-token rate of \autoref{sec:sustained:time}.
The per-user gain grows with history length. Grouped by history length, users with fewer than 100 records gain 0 to 4 points on average over the task adapter, and the eight users with 100 or more gain 17.
The difference between the on-device and the server-trained adapter does not vary with history length, so on-device training reproduces server training for short and long histories alike.

%% file: sections/09-related.tex
\section{Related Work}
\label{sec:related}

\paragraph{Profiling LLMs on mobile devices.}
MELT~\cite{MELT_laskaridis2024} characterized on-device LLM inference across Android, iOS and Jetson, measuring throughput, memory, energy and thermal behavior.
HeteroLLM~\cite{heterollm_chen25} and \citet{scalingLLMNPU_Hao26} profile mobile GPUs and NPUs during inference and schedule work onto the matrix units they find idle.
Thermal throttling of sustained inference has been measured on phones before~\cite{MELT_laskaridis2024,tummalapalli2026edgesustained}, and energy has been reduced through frequency selection~\cite{zou2026enerinfer,zhang2025dvfs} and workload-level knobs such as speculative decoding~\cite{yang2026pelm} for inference, and through frequency selection for fine-tuning on an embedded GPU~\cite{park2026dvfsfinetune}.
Our study asks the same questions of training on a platform that exposes no frequency control to applications.

\paragraph{Training on mobile devices.}
PockEngine~\cite{zhu2023pockengine} derives a training graph at compile time and prunes the backward pass. Sage, Melon and Mandheling~\cite{gim2022sage,wang2022melon,xu2022mandheling} address the memory wall and integer training on mobile hardware.
For LLMs, \citet{mobilemebpsong25} fit a 4B model into an iPhone's memory with lazy weight loading and checkpointing, but measure only individual training steps on the phone, at 256 tokens, and evaluate training quality in a server-side simulation. MeSP~\cite{park2026mesp} halves MeBP's memory, again measuring single steps, on an iPhone~17~Pro with MLX.
MobileFineTuner~\cite{mobilefinetuner2025} trains models up to 1B on Android with a native stack and estimates energy from the platform's power model, FBLayout~\cite{fblayout_tam26} tunes memory layout for both passes on mobile GPUs, and EdgeTune~\cite{wang2026edgetune}, noting that LoRA leaves about two-thirds of the fine-tuning arithmetic in the frozen backbone, removes low-yield adapter paths to reduce it on models of up to 1.1B parameters.
Like server-side breakdowns~\cite{zhu2026lorafusion}, our breakdown separates the adapter from the frozen base, and for a 3B model on a phone it places the adapter at 4--7\% of the step, which bounds what adapter-side optimization can return, and identifies adapter depth as the placement choice that moves compute.
PocketLLM~\cite{pocketllm2024published} and MobiZO~\cite{gao2025mobizo} train on the device without gradients through derivative-free optimization~\cite{malladi2023MeZO}, and MobiLLM and its successor~\cite{mobillm2025,yang2025paemobillm} keep only the frozen forward pass on the device and train a side network on a server. CDCDA-PLM~\cite{mobicdcda2025} fine-tunes a per-user adapter on cloud-generated synthetic data, but runs the training on a workstation GPU and deploys the result to the device. We show that the full backward pass is affordable on the phone at 3B once its kernel is correct.

\paragraph{Quantized Training.}
QLoRA~\cite{dettmers2023qlora} backpropagates through a frozen 4-bit base into 16-bit adapters. Its reference implementation~\cite{bitsandbytesrepo} runs a fused 4-bit kernel in the forward pass for inputs of up to 1,536 tokens and dequantizes the weight to a dense matrix for the backward pass, so the reference implementation itself accelerates only the direction that inference needs.
GPTQ and AWQ~\cite{frantar2022gptq,lin2024awq} quantize for inference and fix the layout that makes the forward contraction cheap.
The runtimes we audited in \S\ref{sec:time:audit} inherit that layout. ggml~\cite{ggml}, the tensor library of llama.cpp, has no mobile GPU kernel for the backward contraction over packed weights and dequantizes the matrix on the CPU once per gradient row instead. Concurrently with our work, ExecuTorch added 4-bit training kernels to its Vulkan and WebGPU backends, including the backward pass of the quantized linear layer and an AdamW optimizer step~\cite{executorchq4gsw}, whereas \citet{executorch2026} validate on-device training only on a CIFAR-10 image classifier running on the CPU.
We measure what such a kernel changes over complete training runs of a 3B-parameter model on a phone, and our audit shows that the gap it closes is common to runtimes designed for inference.
On the server side, \citet{chen2025celora} make the complementary point that freezing the base removes only the products that compute its weight gradients, about a third of the arithmetic, and \citet{sung2022lst} that the backward pass must still run through it.

\paragraph{Personalization and privacy.}
Personalizing an LLM from user history is a growing field~\cite{zhang2025personalizationlargelanguagemodels}. \lamp{}~\cite{lamp2024} and OPPU~\cite{oppu2024} define the tasks and the two-stage adapter recipe we adopt, and \citet{salemi2024ragvsft} compare retrieval and fine-tuning for personalization on servers.
Federated and split learning keep raw data local but send gradients or activations off the device~\cite{fedlearningmcmahan2017,splitlearningvepakomma2018,chen2026flux,geng2026edgeflowertune}, and inputs can be reconstructed from either~\cite{zhu2019dlg,he2019modelinversion,pasquini2021unleashtiger}. Trusted execution on phones~\cite{wang2026tzllm} addresses the inverse problem, protecting a vendor's model weights from the device owner.

%% file: sections/10-discussion.tex
\section{Discussion and Limitations}
\label{sec:discussion}
\paragraph{Cost levers.}
The cost of on-device training is decided in the frozen base, not in the adapter, since every step still runs a forward and a backward pass through the full model and the backward pass over quantized weights takes most of the step.
The obstacle there is software rather than arithmetic, since the two contractions need different kernels, and the runtimes we audited either have no accelerated backward kernel or, in the case of MLX, had one that was never dispatched, except ExecuTorch, which added one concurrently with this work.
Once dispatched, the backward pass runs at the cost its arithmetic predicts, and the MLX defect went unnoticed because no training workload or test exercised that kernel.
What the application controls is the shape of the adapter, where rank changes the cost little while the number of adapted projections, the adapted depth, and the number of passes over the history change it substantially, and no schedule we tested lowers it.

\paragraph{Limitations.}
Our measurements are of one device and one platform, and iOS is the restrictive one, since Android exposes a thermal headroom forecast~\cite{androidthermal}, performance hint sessions~\cite{androidadpf} and foreground services that keep work running in the background~\cite{androidforeground}, so our findings on scheduling and on the thermal signal are findings about iOS.

\paragraph{What runtimes and operating systems should provide.}
A runtime that supports training needs the backward contraction over quantized weights as a kernel.
Because the base is frozen, its layout can be chosen at load time for both contraction directions rather than inherited from the inference format.
An operating system could support on-device training by letting it run while the device is charging and off screen, which removes the battery cost at the price of the 5 to 10\% slowdown of our plugged-in runs.
An adapter runs for hours without a deadline and can save its few megabytes of state at every step, like the photo indexing and backups that a phone already defers to charging periods, yet iOS grants such work budgets of minutes and no GPU~\cite{applebgprocessingtask}.

%% file: sections/11-conclusion.tex
\section{Conclusion}
\label{sec:conclusion}
A phone can personalize its language model by training on the device.
On an iPhone~17~Pro, per-user adapters for a 3B-parameter model at 4-bit precision improve personalization benchmark scores and match adapters trained on a server.
Training time and energy are linear in the number of tokens in the user's history, and a 405-record history costs about half a battery charge.

Once adapters, quantization and checkpointing fit training into memory, its cost lies in the backward pass over the frozen quantized base model, which runs through kernels and weight layouts designed for inference.
In MLX the accelerated kernel for this pass had never been dispatched and was incorrect, and nine of the ten other runtimes we audited lack such a kernel.
Sustained training is further limited by the thermal envelope of the device and by an operating system that restricts GPU training to the foreground.
We repaired the kernel in MLX. The weight layout, the testing of training paths in inference runtimes, and background execution remain open, and they will determine the cost of on-device training on future devices.

\paragraph{Use of AI tools.}
We used Claude (Anthropic) to help write code, run experiments, search related work, and edit the text, and we checked and verified all of its output.

%% file: main.bbl

\begin{thebibliography}{78}


\ifx \showCODEN    \undefined \def \showCODEN     #1{\unskip}     \fi
\ifx \showISBNx    \undefined \def \showISBNx     #1{\unskip}     \fi
\ifx \showISBNxiii \undefined \def \showISBNxiii  #1{\unskip}     \fi
\ifx \showISSN     \undefined \def \showISSN      #1{\unskip}     \fi
\ifx \showLCCN     \undefined \def \showLCCN      #1{\unskip}     \fi
\ifx \shownote     \undefined \def \shownote      #1{#1}          \fi
\ifx \showarticletitle \undefined \def \showarticletitle #1{#1}   \fi
\ifx \showURL      \undefined \def \showURL       {\relax}        \fi
\providecommand\bibfield[2]{#2}
\providecommand\bibinfo[2]{#2}
\providecommand\natexlab[1]{#1}
\providecommand\showeprint[2][]{arXiv:#2}

\bibitem[{Apple Distribution International Limited}(2025)]%
        {eprel17pro}
\bibfield{author}{\bibinfo{person}{{Apple Distribution International
  Limited}}.} \bibinfo{year}{2025}\natexlab{}.
\newblock \bibinfo{title}{Product Information Sheet, model {A3523} ({EPREL}
  registration 2402615)}.
\newblock
  \bibinfo{howpublished}{\url{https://eprel.ec.europa.eu/fiches/smartphonestablets20231669/Fiche_2402615_EN.pdf}}.
\newblock
\newblock
\shownote{European Product Registry for Energy Labelling, pursuant to
  Commission Delegated Regulation (EU) 2023/1669. Accessed 2026-09-11}.


\bibitem[{Apple Inc.}(2025)]%
        {applea19pro}
\bibfield{author}{\bibinfo{person}{{Apple Inc.}}}
  \bibinfo{year}{2025}\natexlab{}.
\newblock \bibinfo{title}{Apple unveils {iPhone} 17 {Pro} and {iPhone} 17 {Pro
  Max}}.
\newblock
  \bibinfo{howpublished}{\url{https://www.apple.com/newsroom/2025/09/apple-unveils-iphone-17-pro-and-iphone-17-pro-max/}}.
\newblock
\newblock
\shownote{Apple Newsroom press release, 9 September 2025}.


\bibitem[{Apple Inc.}(2026a)]%
        {applebggpu}
\bibfield{author}{\bibinfo{person}{{Apple Inc.}}}
  \bibinfo{year}{2026}\natexlab{a}.
\newblock \bibinfo{title}{{BGContinuedProcessingTaskRequest.Resources.gpu}}.
\newblock
  \bibinfo{howpublished}{\url{https://developer.apple.com/documentation/backgroundtasks/bgcontinuedprocessingtaskrequest/resources/gpu}}.
\newblock
\newblock
\shownote{Apple Developer Documentation. Accessed 2026-09-11}.


\bibitem[{Apple Inc.}(2026b)]%
        {applebgprocessingtask}
\bibfield{author}{\bibinfo{person}{{Apple Inc.}}}
  \bibinfo{year}{2026}\natexlab{b}.
\newblock \bibinfo{title}{{BGProcessingTask}}.
\newblock
  \bibinfo{howpublished}{\url{https://developer.apple.com/documentation/backgroundtasks/bgprocessingtask}}.
\newblock
\newblock
\shownote{Apple Developer Documentation. Accessed 2026-09-11}.


\bibitem[{Apple Inc.}(2026c)]%
        {appleincreasedmemory}
\bibfield{author}{\bibinfo{person}{{Apple Inc.}}}
  \bibinfo{year}{2026}\natexlab{c}.
\newblock \bibinfo{title}{{com.apple.developer.kernel.increased-memory-limit}}.
\newblock
  \bibinfo{howpublished}{\url{https://developer.apple.com/documentation/bundleresources/entitlements/com.apple.developer.kernel.increased-memory-limit}}.
\newblock
\newblock
\shownote{Apple Developer Documentation. Accessed 2026-09-11}.


\bibitem[{Apple Inc.}(2026d)]%
        {applesustainedexec}
\bibfield{author}{\bibinfo{person}{{Apple Inc.}}}
  \bibinfo{year}{2026}\natexlab{d}.
\newblock \bibinfo{title}{{com.apple.developer.sustained-execution}}.
\newblock
  \bibinfo{howpublished}{\url{https://developer.apple.com/documentation/bundleresources/entitlements/com.apple.developer.sustained-execution}}.
\newblock
\newblock
\shownote{Apple Developer Documentation. Accessed 2026-09-24}.


\bibitem[{Apple Inc.}(2026e)]%
        {coreml}
\bibfield{author}{\bibinfo{person}{{Apple Inc.}}}
  \bibinfo{year}{2026}\natexlab{e}.
\newblock \bibinfo{title}{Core {ML}}.
\newblock
  \bibinfo{howpublished}{\url{https://developer.apple.com/documentation/coreml}}.
\newblock
\newblock
\shownote{Accessed 2026-09-11}.


\bibitem[Bakouch et~al\mbox{.}(2025)]%
        {smollm3}
\bibfield{author}{\bibinfo{person}{Elie Bakouch}, \bibinfo{person}{Loubna
  Ben~Allal}, \bibinfo{person}{Anton Lozhkov}, \bibinfo{person}{Nouamane Tazi},
  \bibinfo{person}{Lewis Tunstall}, \bibinfo{person}{Carlos~Miguel Patiño},
  \bibinfo{person}{Edward Beeching}, \bibinfo{person}{Aymeric Roucher},
  \bibinfo{person}{Aksel~Joonas Reedi}, \bibinfo{person}{Quentin Gallouédec},
  \bibinfo{person}{Kashif Rasul}, \bibinfo{person}{Nathan Habib},
  \bibinfo{person}{Clémentine Fourrier}, \bibinfo{person}{Hynek Kydlicek},
  \bibinfo{person}{Guilherme Penedo}, \bibinfo{person}{Hugo Larcher},
  \bibinfo{person}{Mathieu Morlon}, \bibinfo{person}{Vaibhav Srivastav},
  \bibinfo{person}{Joshua Lochner}, \bibinfo{person}{Xuan-Son Nguyen},
  \bibinfo{person}{Colin Raffel}, \bibinfo{person}{Leandro von Werra}, {and}
  \bibinfo{person}{Thomas Wolf}.} \bibinfo{year}{2025}\natexlab{}.
\newblock \bibinfo{title}{{SmolLM3: smol, multilingual, long-context
  reasoner}}.
\newblock \bibinfo{howpublished}{\url{https://huggingface.co/blog/smollm3}}.
\newblock


\bibitem[{bitsandbytes Foundation}(2026)]%
        {bitsandbytesrepo}
\bibfield{author}{\bibinfo{person}{{bitsandbytes Foundation}}.}
  \bibinfo{year}{2026}\natexlab{}.
\newblock \bibinfo{title}{bitsandbytes}.
\newblock
  \bibinfo{howpublished}{\url{https://github.com/bitsandbytes-foundation/bitsandbytes}}.
\newblock
\newblock
\shownote{Accessed 2026-09-11}.


\bibitem[Chen et~al\mbox{.}(2026)]%
        {chen2026flux}
\bibfield{author}{\bibinfo{person}{Fahao Chen}, \bibinfo{person}{Jie Wan},
  \bibinfo{person}{Peng Li}, \bibinfo{person}{Zhou Su}, {and}
  \bibinfo{person}{Dongxiao Yu}.} \bibinfo{year}{2026}\natexlab{}.
\newblock \showarticletitle{Federated Fine-Tuning of Sparsely-Activated Large
  Language Models on Resource-Constrained Devices}. In
  \bibinfo{booktitle}{\emph{Proceedings of the 21st European Conference on
  Computer Systems (EuroSys '26)}}. \bibinfo{publisher}{{ACM}}.
\newblock
\newblock
\shownote{System named FLUX; arXiv:2508.19078}.


\bibitem[Chen et~al\mbox{.}(2025b)]%
        {chen2025celora}
\bibfield{author}{\bibinfo{person}{Guanduo Chen}, \bibinfo{person}{Yutong He},
  \bibinfo{person}{Yipeng Hu}, \bibinfo{person}{Kun Yuan}, {and}
  \bibinfo{person}{Binhang Yuan}.} \bibinfo{year}{2025}\natexlab{b}.
\newblock \bibinfo{title}{{CE-LoRA}: Computation-Efficient {LoRA} Fine-Tuning
  for Language Models}.
\newblock
\showeprint[arxiv]{2502.01378}~[cs.LG]
\urldef\tempurl%
\url{https://arxiv.org/abs/2502.01378}
\showURL{%
\tempurl}


\bibitem[Chen et~al\mbox{.}(2025a)]%
        {heterollm_chen25}
\bibfield{author}{\bibinfo{person}{Le Chen}, \bibinfo{person}{Dahu Feng},
  \bibinfo{person}{Erhu Feng}, \bibinfo{person}{Yingrui Wang},
  \bibinfo{person}{Rong Zhao}, \bibinfo{person}{Yubin Xia},
  \bibinfo{person}{Pinjie Xu}, {and} \bibinfo{person}{Haibo Chen}.}
  \bibinfo{year}{2025}\natexlab{a}.
\newblock \showarticletitle{Characterizing Mobile {SoC} for Accelerating
  Heterogeneous {LLM} Inference}. In \bibinfo{booktitle}{\emph{Proceedings of
  the ACM SIGOPS 31st Symposium on Operating Systems Principles (SOSP '25)}}.
  \bibinfo{publisher}{{ACM}}, \bibinfo{pages}{359--374}.
\newblock
\href{https://doi.org/10.1145/3731569.3764808}{doi:\nolinkurl{10.1145/3731569.3764808}}


\bibitem[Chen et~al\mbox{.}(2018)]%
        {tvm2018}
\bibfield{author}{\bibinfo{person}{Tianqi Chen}, \bibinfo{person}{Thierry
  Moreau}, \bibinfo{person}{Ziheng Jiang}, \bibinfo{person}{Lianmin Zheng},
  \bibinfo{person}{Eddie Yan}, \bibinfo{person}{Haichen Shen},
  \bibinfo{person}{Meghan Cowan}, \bibinfo{person}{Leyuan Wang},
  \bibinfo{person}{Yuwei Hu}, \bibinfo{person}{Luis Ceze},
  \bibinfo{person}{Carlos Guestrin}, {and} \bibinfo{person}{Arvind
  Krishnamurthy}.} \bibinfo{year}{2018}\natexlab{}.
\newblock \showarticletitle{{TVM}: An Automated End-to-End Optimizing Compiler
  for Deep Learning}. In \bibinfo{booktitle}{\emph{13th USENIX Symposium on
  Operating Systems Design and Implementation (OSDI 18)}}.
  \bibinfo{publisher}{USENIX Association}, \bibinfo{pages}{578--594}.
\newblock
\urldef\tempurl%
\url{https://www.usenix.org/conference/osdi18/presentation/chen}
\showURL{%
\tempurl}


\bibitem[Chen et~al\mbox{.}(2016)]%
        {chen2016gradientcheckpointing}
\bibfield{author}{\bibinfo{person}{Tianqi Chen}, \bibinfo{person}{Bing Xu},
  \bibinfo{person}{Chiyuan Zhang}, {and} \bibinfo{person}{Carlos Guestrin}.}
  \bibinfo{year}{2016}\natexlab{}.
\newblock \showarticletitle{Training Deep Nets with Sublinear Memory Cost}.
\newblock \bibinfo{journal}{\emph{CoRR}}  \bibinfo{volume}{abs/1604.06174}
  (\bibinfo{year}{2016}).
\newblock
\showeprint[arXiv]{1604.06174}
\urldef\tempurl%
\url{http://arxiv.org/abs/1604.06174}
\showURL{%
\tempurl}


\bibitem[Chen et~al\mbox{.}(2025c)]%
        {chen2025confidant}
\bibfield{author}{\bibinfo{person}{Yuhao Chen}, \bibinfo{person}{Yuxuan Yan},
  \bibinfo{person}{Shuowei Ge}, \bibinfo{person}{Yuyang Qin},
  \bibinfo{person}{Yue Zheng}, \bibinfo{person}{Qianqian Yang},
  \bibinfo{person}{Shibo He}, \bibinfo{person}{Zhiguo Shi},
  \bibinfo{person}{Jiming Chen}, {and} \bibinfo{person}{Yuanchao Shu}.}
  \bibinfo{year}{2025}\natexlab{c}.
\newblock \showarticletitle{Confidant: Customizing Transformer-based {LLMs} via
  Collaborative Training on Mobile Devices}. In
  \bibinfo{booktitle}{\emph{Proceedings of the 31st Annual International
  Conference on Mobile Computing and Networking (MobiCom '25)}}.
  \bibinfo{pages}{483--497}.
\newblock
\href{https://doi.org/10.1145/3680207.3723487}{doi:\nolinkurl{10.1145/3680207.3723487}}


\bibitem[Dettmers et~al\mbox{.}(2023)]%
        {dettmers2023qlora}
\bibfield{author}{\bibinfo{person}{Tim Dettmers}, \bibinfo{person}{Artidoro
  Pagnoni}, \bibinfo{person}{Ari Holtzman}, {and} \bibinfo{person}{Luke
  Zettlemoyer}.} \bibinfo{year}{2023}\natexlab{}.
\newblock \showarticletitle{{QLoRA}: Efficient Finetuning of Quantized {LLM}s}.
  In \bibinfo{booktitle}{\emph{Advances in Neural Information Processing
  Systems 36 (NeurIPS 2023)}}.
\newblock
\href{https://doi.org/10.52202/075280-0441}{doi:\nolinkurl{10.52202/075280-0441}}
\showeprint[arxiv]{2305.14314}~[cs.LG]


\bibitem[Frantar et~al\mbox{.}(2022)]%
        {frantar2022gptq}
\bibfield{author}{\bibinfo{person}{Elias Frantar}, \bibinfo{person}{Saleh
  Ashkboos}, \bibinfo{person}{Torsten Hoefler}, {and} \bibinfo{person}{Dan
  Alistarh}.} \bibinfo{year}{2022}\natexlab{}.
\newblock \bibinfo{title}{{GPTQ}: Accurate Post-Training Quantization for
  Generative Pre-trained Transformers}.
\newblock
\showeprint[arxiv]{2210.17323}~[cs.LG]
\urldef\tempurl%
\url{https://arxiv.org/abs/2210.17323}
\showURL{%
\tempurl}


\bibitem[Gao et~al\mbox{.}(2025)]%
        {gao2025mobizo}
\bibfield{author}{\bibinfo{person}{Lei Gao}, \bibinfo{person}{Amir Ziashahabi},
  \bibinfo{person}{Yue Niu}, \bibinfo{person}{Salman Avestimehr}, {and}
  \bibinfo{person}{Murali Annavaram}.} \bibinfo{year}{2025}\natexlab{}.
\newblock \showarticletitle{{MobiZO}: Enabling Efficient {LLM} Fine-Tuning at
  the Edge via Inference Engines}. In \bibinfo{booktitle}{\emph{Proceedings of
  the 2025 Conference on Empirical Methods in Natural Language Processing}}.
  \bibinfo{pages}{20217--20234}.
\newblock
\href{https://doi.org/10.18653/v1/2025.emnlp-main.1022}{doi:\nolinkurl{10.18653/v1/2025.emnlp-main.1022}}


\bibitem[{Gemini Team}(2023)]%
        {gemini2023}
\bibfield{author}{\bibinfo{person}{{Gemini Team}}.}
  \bibinfo{year}{2023}\natexlab{}.
\newblock \bibinfo{title}{Gemini: A Family of Highly Capable Multimodal
  Models}.
\newblock
\showeprint[arxiv]{2312.11805}~[cs.CL]
\urldef\tempurl%
\url{https://arxiv.org/abs/2312.11805}
\showURL{%
\tempurl}


\bibitem[Geng et~al\mbox{.}(2026)]%
        {geng2026edgeflowertune}
\bibfield{author}{\bibinfo{person}{Jiaxiang Geng}, \bibinfo{person}{Yiyi Lu},
  \bibinfo{person}{Lunyu Zhao}, \bibinfo{person}{Yan Gao},
  \bibinfo{person}{Nicholas~D. Lane}, {and} \bibinfo{person}{Bing Luo}.}
  \bibinfo{year}{2026}\natexlab{}.
\newblock \bibinfo{title}{{EdgeFlowerTune}: Evaluating Federated {LLM}
  Fine-Tuning Under Realistic Edge System Constraints}.
\newblock
\showeprint[arxiv]{2605.08636}~[cs.LG]
\urldef\tempurl%
\url{https://arxiv.org/abs/2605.08636}
\showURL{%
\tempurl}


\bibitem[Geng et~al\mbox{.}(2025)]%
        {mobilefinetuner2025}
\bibfield{author}{\bibinfo{person}{Jiaxiang Geng}, \bibinfo{person}{Lunyu
  Zhao}, \bibinfo{person}{Yiyi Lu}, {and} \bibinfo{person}{Bing Luo}.}
  \bibinfo{year}{2025}\natexlab{}.
\newblock \showarticletitle{{MobileFineTuner}: A Mobile-Native Framework for
  On-Device {LLM} Fine-Tuning in Real-World Embedded {AI} Applications}.
\newblock \bibinfo{journal}{\emph{arXiv preprint arXiv:2512.08211}}
  (\bibinfo{year}{2025}).
\newblock


\bibitem[Gerganov et~al\mbox{.}(2022)]%
        {ggml}
\bibfield{author}{\bibinfo{person}{Georgi Gerganov} {et~al\mbox{.}}}
  \bibinfo{year}{2022}\natexlab{}.
\newblock \bibinfo{title}{{ggml}: Tensor library for machine learning}.
\newblock \bibinfo{howpublished}{\url{https://github.com/ggml-org/ggml}}.
\newblock


\bibitem[Gerganov et~al\mbox{.}(2023)]%
        {llamacpp2023}
\bibfield{author}{\bibinfo{person}{Georgi Gerganov} {et~al\mbox{.}}}
  \bibinfo{year}{2023}\natexlab{}.
\newblock \bibinfo{title}{{llama.cpp}}.
\newblock
\urldef\tempurl%
\url{https://github.com/ggml-org/llama.cpp}
\showURL{%
\tempurl}


\bibitem[Gim and Ko(2022)]%
        {gim2022sage}
\bibfield{author}{\bibinfo{person}{In Gim} {and} \bibinfo{person}{JeongGil
  Ko}.} \bibinfo{year}{2022}\natexlab{}.
\newblock \showarticletitle{Memory-efficient {DNN} training on mobile devices}.
  In \bibinfo{booktitle}{\emph{Proceedings of the 20th Annual International
  Conference on Mobile Systems, Applications and Services (MobiSys '22)}}.
  \bibinfo{publisher}{{ACM}}, \bibinfo{pages}{464--476}.
\newblock
\href{https://doi.org/10.1145/3498361.3539765}{doi:\nolinkurl{10.1145/3498361.3539765}}
\newblock
\shownote{System named Sage}.


\bibitem[{Google}(2026a)]%
        {androidforeground}
\bibfield{author}{\bibinfo{person}{{Google}}.}
  \bibinfo{year}{2026}\natexlab{a}.
\newblock \bibinfo{title}{Foreground services overview}.
\newblock \bibinfo{howpublished}{Android Developers documentation}.
\newblock
\urldef\tempurl%
\url{https://developer.android.com/develop/background-work/services/fgs}
\showURL{%
\tempurl}
\newblock
\shownote{Accessed September 2026}.


\bibitem[{Google}(2026b)]%
        {androidadpf}
\bibfield{author}{\bibinfo{person}{{Google}}.}
  \bibinfo{year}{2026}\natexlab{b}.
\newblock \bibinfo{title}{Performance Hint {API}}.
\newblock \bibinfo{howpublished}{Android Developers documentation}.
\newblock
\urldef\tempurl%
\url{https://developer.android.com/games/optimize/adpf/performance-hint-api}
\showURL{%
\tempurl}
\newblock
\shownote{Accessed September 2026}.


\bibitem[{Google}(2026c)]%
        {androidthermal}
\bibfield{author}{\bibinfo{person}{{Google}}.}
  \bibinfo{year}{2026}\natexlab{c}.
\newblock \bibinfo{title}{{PowerManager.getThermalHeadroom}}.
\newblock \bibinfo{howpublished}{Android Developers documentation}.
\newblock
\urldef\tempurl%
\url{https://developer.android.com/reference/android/os/PowerManager}
\showURL{%
\tempurl}
\newblock
\shownote{Accessed September 2026}.


\bibitem[{Google AI Edge}(2024)]%
        {litert}
\bibfield{author}{\bibinfo{person}{{Google AI Edge}}.}
  \bibinfo{year}{2024}\natexlab{}.
\newblock \bibinfo{title}{{LiteRT}: Google's On-Device Runtime for
  High-Performance {ML} and {GenAI}}.
\newblock
  \bibinfo{howpublished}{\url{https://github.com/google-ai-edge/LiteRT}}.
\newblock


\bibitem[Hannun et~al\mbox{.}(2023)]%
        {mlx2023}
\bibfield{author}{\bibinfo{person}{Awni Hannun}, \bibinfo{person}{Jagrit
  Digani}, \bibinfo{person}{Angelos Katharopoulos}, {and}
  \bibinfo{person}{Ronan Collobert}.} \bibinfo{year}{2023}\natexlab{}.
\newblock \bibinfo{title}{{MLX}: Efficient and flexible machine learning on
  Apple silicon}.
\newblock
\urldef\tempurl%
\url{https://github.com/ml-explore}
\showURL{%
\tempurl}


\bibitem[Hao et~al\mbox{.}(2026)]%
        {scalingLLMNPU_Hao26}
\bibfield{author}{\bibinfo{person}{Zixu Hao}, \bibinfo{person}{Jianyu Wei},
  \bibinfo{person}{Tuowei Wang}, \bibinfo{person}{Minxing Huang},
  \bibinfo{person}{Huiqiang Jiang}, \bibinfo{person}{Shiqi Jiang},
  \bibinfo{person}{Ting Cao}, {and} \bibinfo{person}{Ju Ren}.}
  \bibinfo{year}{2026}\natexlab{}.
\newblock \showarticletitle{Scaling {LLM} Test-Time Compute with Mobile {NPU}
  on Smartphones}. In \bibinfo{booktitle}{\emph{Proceedings of the 21st
  European Conference on Computer Systems, EuroSys 2026, McEwan Hall/The
  University of Edinburgh, Edinburgh, Scotland, UK, April 27-30, 2026}},
  \bibfield{editor}{\bibinfo{person}{Antonio Barbalace}, \bibinfo{person}{Luo
  Mai}, \bibinfo{person}{Roxana Geambasu}, {and} \bibinfo{person}{Peter~R.
  Pietzuch}} (Eds.). \bibinfo{publisher}{{ACM}}, \bibinfo{pages}{2157--2172}.
\newblock
\href{https://doi.org/10.1145/3767295.3769382}{doi:\nolinkurl{10.1145/3767295.3769382}}


\bibitem[He et~al\mbox{.}(2019)]%
        {he2019modelinversion}
\bibfield{author}{\bibinfo{person}{Zecheng He}, \bibinfo{person}{Tianwei
  Zhang}, {and} \bibinfo{person}{Ruby~B. Lee}.}
  \bibinfo{year}{2019}\natexlab{}.
\newblock \showarticletitle{Model inversion attacks against collaborative
  inference}. In \bibinfo{booktitle}{\emph{Proceedings of the 35th Annual
  Computer Security Applications Conference}} (San Juan, Puerto Rico, USA)
  \emph{(\bibinfo{series}{ACSAC '19})}. \bibinfo{publisher}{Association for
  Computing Machinery}, \bibinfo{address}{New York, NY, USA},
  \bibinfo{pages}{148–162}.
\newblock
\showISBNx{9781450376280}
\href{https://doi.org/10.1145/3359789.3359824}{doi:\nolinkurl{10.1145/3359789.3359824}}


\bibitem[Hu et~al\mbox{.}(2021)]%
        {hu2021loralowrankadaptationlarge}
\bibfield{author}{\bibinfo{person}{Edward~J. Hu}, \bibinfo{person}{Yelong
  Shen}, \bibinfo{person}{Phillip Wallis}, \bibinfo{person}{Zeyuan Allen-Zhu},
  \bibinfo{person}{Yuanzhi Li}, \bibinfo{person}{Shean Wang},
  \bibinfo{person}{Lu Wang}, {and} \bibinfo{person}{Weizhu Chen}.}
  \bibinfo{year}{2021}\natexlab{}.
\newblock \bibinfo{title}{LoRA: Low-Rank Adaptation of Large Language Models}.
\newblock
\showeprint[arxiv]{2106.09685}~[cs.CL]
\urldef\tempurl%
\url{https://arxiv.org/abs/2106.09685}
\showURL{%
\tempurl}


\bibitem[Jiang et~al\mbox{.}(2020)]%
        {mnn2020}
\bibfield{author}{\bibinfo{person}{Xiaotang Jiang}, \bibinfo{person}{Huan
  Wang}, \bibinfo{person}{Yiliu Chen}, \bibinfo{person}{Ziqi Wu},
  \bibinfo{person}{Lichuan Wang}, \bibinfo{person}{Bin Zou},
  \bibinfo{person}{Yafeng Yang}, \bibinfo{person}{Zongyang Cui},
  \bibinfo{person}{Yu Cai}, \bibinfo{person}{Tianhang Yu},
  \bibinfo{person}{Chengfei Lv}, {and} \bibinfo{person}{Zhihua Wu}.}
  \bibinfo{year}{2020}\natexlab{}.
\newblock \showarticletitle{MNN: A Universal and Efficient Inference Engine}.
  In \bibinfo{booktitle}{\emph{MLSys}}.
\newblock


\bibitem[Laskaridis et~al\mbox{.}(2024)]%
        {MELT_laskaridis2024}
\bibfield{author}{\bibinfo{person}{Stefanos Laskaridis},
  \bibinfo{person}{Kleomenis Katevas}, \bibinfo{person}{Lorenzo Minto}, {and}
  \bibinfo{person}{Hamed Haddadi}.} \bibinfo{year}{2024}\natexlab{}.
\newblock \showarticletitle{MELTing Point: Mobile Evaluation of Language
  Transformers}. In \bibinfo{booktitle}{\emph{Proceedings of the 30th Annual
  International Conference on Mobile Computing and Networking, {ACM} MobiCom
  2024, Washington D.C., DC, USA, November 18-22, 2024}},
  \bibfield{editor}{\bibinfo{person}{Weisong Shi}, \bibinfo{person}{Deepak
  Ganesan}, {and} \bibinfo{person}{Nicholas~D. Lane}} (Eds.).
  \bibinfo{publisher}{{ACM}}, \bibinfo{pages}{890--907}.
\newblock
\href{https://doi.org/10.1145/3636534.3690668}{doi:\nolinkurl{10.1145/3636534.3690668}}


\bibitem[Li et~al\mbox{.}(2025a)]%
        {applefm2025}
\bibfield{author}{\bibinfo{person}{Ethan Li}, \bibinfo{person}{Anders
  Boesen~Lindbo Larsen}, \bibinfo{person}{Chen Zhang}, {et~al\mbox{.}}}
  \bibinfo{year}{2025}\natexlab{a}.
\newblock \bibinfo{title}{Apple Intelligence Foundation Language Models: Tech
  Report 2025}.
\newblock
\showeprint[arxiv]{2507.13575}~[cs.LG]
\urldef\tempurl%
\url{https://arxiv.org/abs/2507.13575}
\showURL{%
\tempurl}


\bibitem[Li et~al\mbox{.}(2025b)]%
        {mobillm2025}
\bibfield{author}{\bibinfo{person}{Liang Li}, \bibinfo{person}{Xingke Yang},
  \bibinfo{person}{Wen Wu}, \bibinfo{person}{Hao Wang},
  \bibinfo{person}{Tomoaki Ohtsuki}, \bibinfo{person}{Xin Fu},
  \bibinfo{person}{Miao Pan}, {and} \bibinfo{person}{Xuemin Shen}.}
  \bibinfo{year}{2025}\natexlab{b}.
\newblock \showarticletitle{MobiLLM: Enabling LLM Fine-Tuning on the Mobile
  Device via Server Assisted Side Tuning}.
\newblock \bibinfo{journal}{\emph{arXiv preprint arXiv:2502.20421}}
  (\bibinfo{year}{2025}).
\newblock


\bibitem[Lin(2004)]%
        {lin2004rouge}
\bibfield{author}{\bibinfo{person}{Chin-Yew Lin}.}
  \bibinfo{year}{2004}\natexlab{}.
\newblock \showarticletitle{{ROUGE}: A Package for Automatic Evaluation of
  Summaries}. In \bibinfo{booktitle}{\emph{Text Summarization Branches Out}}.
  \bibinfo{publisher}{Association for Computational Linguistics},
  \bibinfo{address}{Barcelona, Spain}, \bibinfo{pages}{74--81}.
\newblock
\urldef\tempurl%
\url{https://aclanthology.org/W04-1013}
\showURL{%
\tempurl}


\bibitem[Lin et~al\mbox{.}(2024)]%
        {lin2024awq}
\bibfield{author}{\bibinfo{person}{Ji Lin}, \bibinfo{person}{Jiaming Tang},
  \bibinfo{person}{Haotian Tang}, \bibinfo{person}{Shang Yang},
  \bibinfo{person}{Wei-Ming Chen}, \bibinfo{person}{Wei-Chen Wang},
  \bibinfo{person}{Guangxuan Xiao}, \bibinfo{person}{Xingyu Dang},
  \bibinfo{person}{Chuang Gan}, {and} \bibinfo{person}{Song Han}.}
  \bibinfo{year}{2024}\natexlab{}.
\newblock \showarticletitle{{AWQ}: Activation-aware Weight Quantization for
  On-Device {LLM} Compression and Acceleration}. In
  \bibinfo{booktitle}{\emph{Proceedings of Machine Learning and Systems 6
  (MLSys 2024)}}.
\newblock
\showeprint[arxiv]{2306.00978}~[cs.CL]
\urldef\tempurl%
\url{https://proceedings.mlsys.org/paper_files/paper/2024/hash/42a452cbafa9dd64e9ba4aa95cc1ef21-Abstract-Conference.html}
\showURL{%
\tempurl}


\bibitem[Loshchilov and Hutter(2019)]%
        {loshchilov2019adamw}
\bibfield{author}{\bibinfo{person}{Ilya Loshchilov} {and}
  \bibinfo{person}{Frank Hutter}.} \bibinfo{year}{2019}\natexlab{}.
\newblock \showarticletitle{Decoupled Weight Decay Regularization}. In
  \bibinfo{booktitle}{\emph{International Conference on Learning
  Representations (ICLR)}}.
\newblock
\urldef\tempurl%
\url{https://openreview.net/forum?id=Bkg6RiCqY7}
\showURL{%
\tempurl}


\bibitem[Malladi et~al\mbox{.}(2023)]%
        {malladi2023MeZO}
\bibfield{author}{\bibinfo{person}{Sadhika Malladi}, \bibinfo{person}{Tianyu
  Gao}, \bibinfo{person}{Eshaan Nichani}, \bibinfo{person}{Alex Damian},
  \bibinfo{person}{Jason~D Lee}, \bibinfo{person}{Danqi Chen}, {and}
  \bibinfo{person}{Sanjeev Arora}.} \bibinfo{year}{2023}\natexlab{}.
\newblock \showarticletitle{Fine-Tuning Language Models with Just Forward
  Passes}. In \bibinfo{booktitle}{\emph{Advances in Neural Information
  Processing Systems}}, \bibfield{editor}{\bibinfo{person}{A.~Oh},
  \bibinfo{person}{T.~Naumann}, \bibinfo{person}{A.~Globerson},
  \bibinfo{person}{K.~Saenko}, \bibinfo{person}{M.~Hardt}, {and}
  \bibinfo{person}{S.~Levine}} (Eds.), Vol.~\bibinfo{volume}{36}.
  \bibinfo{publisher}{Curran Associates, Inc.}, \bibinfo{pages}{53038--53075}.
\newblock
\href{https://doi.org/10.52202/075280-2308}{doi:\nolinkurl{10.52202/075280-2308}}


\bibitem[McMahan et~al\mbox{.}(2017)]%
        {fedlearningmcmahan2017}
\bibfield{author}{\bibinfo{person}{Brendan McMahan}, \bibinfo{person}{Eider
  Moore}, \bibinfo{person}{Daniel Ramage}, \bibinfo{person}{Seth Hampson},
  {and} \bibinfo{person}{Blaise~Ag{\"{u}}era y Arcas}.}
  \bibinfo{year}{2017}\natexlab{}.
\newblock \showarticletitle{Communication-Efficient Learning of Deep Networks
  from Decentralized Data}. In \bibinfo{booktitle}{\emph{Proceedings of the
  20th International Conference on Artificial Intelligence and Statistics,
  {AISTATS} 2017, 20-22 April 2017, Fort Lauderdale, FL, {USA}}}
  \emph{(\bibinfo{series}{Proceedings of Machine Learning Research},
  Vol.~\bibinfo{volume}{54})}, \bibfield{editor}{\bibinfo{person}{Aarti Singh}
  {and} \bibinfo{person}{Xiaojin~(Jerry) Zhu}} (Eds.).
  \bibinfo{publisher}{{PMLR}}, \bibinfo{pages}{1273--1282}.
\newblock
\urldef\tempurl%
\url{http://proceedings.mlr.press/v54/mcmahan17a.html}
\showURL{%
\tempurl}


\bibitem[{MLC team}(2025)]%
        {mlcllm}
\bibfield{author}{\bibinfo{person}{{MLC team}}.}
  \bibinfo{year}{2023--2025}\natexlab{}.
\newblock \bibinfo{booktitle}{\emph{{MLC-LLM}}}.
\newblock
\urldef\tempurl%
\url{https://github.com/mlc-ai/mlc-llm}
\showURL{%
\tempurl}


\bibitem[Nachin et~al\mbox{.}(2026)]%
        {executorch2026}
\bibfield{author}{\bibinfo{person}{Mergen Nachin}, \bibinfo{person}{Digant
  Desai}, \bibinfo{person}{Sicheng~Stephen Jia}, \bibinfo{person}{Chen Lai},
  \bibinfo{person}{Mengwei Liu}, \bibinfo{person}{Jacob Szwejbka},
  \bibinfo{person}{Raziel Alvarez}, \bibinfo{person}{RJ Ascani},
  \bibinfo{person}{Dave Bort}, \bibinfo{person}{Manuel Candales},
  {et~al\mbox{.}}} \bibinfo{year}{2026}\natexlab{}.
\newblock \showarticletitle{{ExecuTorch} - A Unified {PyTorch} Solution to Run
  {AI} Models On-Device}.
\newblock \bibinfo{journal}{\emph{arXiv preprint arXiv:2605.08195}}
  (\bibinfo{year}{2026}).
\newblock
\urldef\tempurl%
\url{https://github.com/pytorch/executorch}
\showURL{%
\tempurl}


\bibitem[{ONNX Runtime developers}(2021)]%
        {onnxruntime}
\bibfield{author}{\bibinfo{person}{{ONNX Runtime developers}}.}
  \bibinfo{year}{2021}\natexlab{}.
\newblock \bibinfo{title}{{ONNX} Runtime}.
\newblock \bibinfo{howpublished}{\url{https://onnxruntime.ai/}}.
\newblock
\newblock
\shownote{Version: 1.x}.


\bibitem[Park et~al\mbox{.}(2026a)]%
        {park2026mesp}
\bibfield{author}{\bibinfo{person}{JuneYoung Park}, \bibinfo{person}{Yuri
  Hong}, \bibinfo{person}{Seongwan Kim}, {and} \bibinfo{person}{Jaeho Lee}.}
  \bibinfo{year}{2026}\natexlab{a}.
\newblock \showarticletitle{Memory-Efficient Structured Backpropagation for
  On-Device {LLM} Fine-Tuning}. In \bibinfo{booktitle}{\emph{Proceedings of the
  64th Annual Meeting of the Association for Computational Linguistics (Volume
  6: Industry Track)}}. \bibinfo{pages}{906--916}.
\newblock
\href{https://doi.org/10.18653/v1/2026.acl-industry.62}{doi:\nolinkurl{10.18653/v1/2026.acl-industry.62}}


\bibitem[Park et~al\mbox{.}(2026b)]%
        {park2026dvfsfinetune}
\bibfield{author}{\bibinfo{person}{Jurn-Gyu Park}, \bibinfo{person}{Sanzhar
  Zholdybayev}, \bibinfo{person}{Aidar Amangeldi}, {and} \bibinfo{person}{Ademi
  Zhanuzakova}.} \bibinfo{year}{2026}\natexlab{b}.
\newblock \bibinfo{title}{Energy-Efficient {GPU} {DVFS} for Fine-Tuning of
  {SLM}s on Resource-constrained Embedded Devices}.
\newblock
\showeprint[arxiv]{2607.05933}~[cs.LG]
\urldef\tempurl%
\url{https://arxiv.org/abs/2607.05933}
\showURL{%
\tempurl}


\bibitem[Pasquini et~al\mbox{.}(2021)]%
        {pasquini2021unleashtiger}
\bibfield{author}{\bibinfo{person}{Dario Pasquini}, \bibinfo{person}{Giuseppe
  Ateniese}, {and} \bibinfo{person}{Massimo Bernaschi}.}
  \bibinfo{year}{2021}\natexlab{}.
\newblock \showarticletitle{Unleashing the Tiger: Inference Attacks on Split
  Learning}. In \bibinfo{booktitle}{\emph{Proceedings of the 2021 ACM SIGSAC
  Conference on Computer and Communications Security}} (Virtual Event, Republic
  of Korea) \emph{(\bibinfo{series}{CCS '21})}. \bibinfo{publisher}{Association
  for Computing Machinery}, \bibinfo{address}{New York, NY, USA},
  \bibinfo{pages}{2113–2129}.
\newblock
\showISBNx{9781450384544}
\href{https://doi.org/10.1145/3460120.3485259}{doi:\nolinkurl{10.1145/3460120.3485259}}


\bibitem[Peng et~al\mbox{.}(2024)]%
        {pocketllm2024published}
\bibfield{author}{\bibinfo{person}{Dan Peng}, \bibinfo{person}{Zhihui Fu},
  {and} \bibinfo{person}{Jun Wang}.} \bibinfo{year}{2024}\natexlab{}.
\newblock \showarticletitle{{PocketLLM}: Enabling On-Device Fine-Tuning for
  Personalized {LLM}s}. In \bibinfo{booktitle}{\emph{Proceedings of the Fifth
  Workshop on Privacy in Natural Language Processing (PrivateNLP)}}.
  \bibinfo{publisher}{Association for Computational Linguistics},
  \bibinfo{address}{Bangkok, Thailand}, \bibinfo{pages}{91--96}.
\newblock
\urldef\tempurl%
\url{https://aclanthology.org/2024.privatenlp-1.10/}
\showURL{%
\tempurl}


\bibitem[{PyTorch Contributors}(2026)]%
        {executorchq4gsw}
\bibfield{author}{\bibinfo{person}{{PyTorch Contributors}}.}
  \bibinfo{year}{2026}\natexlab{}.
\newblock \bibinfo{title}{{ExecuTorch} pull requests 20943 and 20931}.
\newblock
  \bibinfo{howpublished}{\url{https://github.com/pytorch/executorch/pull/20943}}.
\newblock
\newblock
\shownote{Add the \texttt{linear\_q4gsw\_backward} kernel to the Vulkan backend
  (20943) and the \texttt{q4gsw} training kernels to the WebGPU backend
  (20931). Both merged 21 July 2026}.


\bibitem[{Qwen Team}(2025)]%
        {qwen3}
\bibfield{author}{\bibinfo{person}{{Qwen Team}}.}
  \bibinfo{year}{2025}\natexlab{}.
\newblock \bibinfo{title}{Qwen3 Technical Report}.
\newblock
\showeprint[arxiv]{2505.09388}~[cs.CL]
\urldef\tempurl%
\url{https://arxiv.org/abs/2505.09388}
\showURL{%
\tempurl}


\bibitem[Robertson and Zaragoza(2009)]%
        {robertson2009bm25}
\bibfield{author}{\bibinfo{person}{Stephen Robertson} {and}
  \bibinfo{person}{Hugo Zaragoza}.} \bibinfo{year}{2009}\natexlab{}.
\newblock \showarticletitle{The Probabilistic Relevance Framework: {BM25} and
  Beyond}.
\newblock \bibinfo{journal}{\emph{Foundations and Trends in Information
  Retrieval}} \bibinfo{volume}{3}, \bibinfo{number}{4} (\bibinfo{year}{2009}),
  \bibinfo{pages}{333--389}.
\newblock
\href{https://doi.org/10.1561/1500000019}{doi:\nolinkurl{10.1561/1500000019}}


\bibitem[R{\"u}ckl{\'e} et~al\mbox{.}(2021)]%
        {ruckle2021adapterdrop}
\bibfield{author}{\bibinfo{person}{Andreas R{\"u}ckl{\'e}},
  \bibinfo{person}{Gregor Geigle}, \bibinfo{person}{Max Glockner},
  \bibinfo{person}{Tilman Beck}, \bibinfo{person}{Jonas Pfeiffer},
  \bibinfo{person}{Nils Reimers}, {and} \bibinfo{person}{Iryna Gurevych}.}
  \bibinfo{year}{2021}\natexlab{}.
\newblock \showarticletitle{{AdapterDrop}: On the Efficiency of Adapters in
  Transformers}. In \bibinfo{booktitle}{\emph{Proceedings of the 2021
  Conference on Empirical Methods in Natural Language Processing}}.
  \bibinfo{pages}{7930--7946}.
\newblock
\href{https://doi.org/10.18653/v1/2021.emnlp-main.626}{doi:\nolinkurl{10.18653/v1/2021.emnlp-main.626}}


\bibitem[Salemi et~al\mbox{.}(2024a)]%
        {salemi2024optimizingpersonalization}
\bibfield{author}{\bibinfo{person}{Alireza Salemi}, \bibinfo{person}{Surya
  Kallumadi}, {and} \bibinfo{person}{Hamed Zamani}.}
  \bibinfo{year}{2024}\natexlab{a}.
\newblock \showarticletitle{Optimization Methods for Personalizing Large
  Language Models through Retrieval Augmentation}. In
  \bibinfo{booktitle}{\emph{Proceedings of the 47th International ACM SIGIR
  Conference on Research and Development in Information Retrieval}} (Washington
  DC, USA) \emph{(\bibinfo{series}{SIGIR '24})}.
  \bibinfo{publisher}{Association for Computing Machinery},
  \bibinfo{address}{New York, NY, USA}, \bibinfo{pages}{752–762}.
\newblock
\showISBNx{9798400704314}
\href{https://doi.org/10.1145/3626772.3657783}{doi:\nolinkurl{10.1145/3626772.3657783}}


\bibitem[Salemi et~al\mbox{.}(2024b)]%
        {lamp2024}
\bibfield{author}{\bibinfo{person}{Alireza Salemi}, \bibinfo{person}{Sheshera
  Mysore}, \bibinfo{person}{Michael Bendersky}, {and} \bibinfo{person}{Hamed
  Zamani}.} \bibinfo{year}{2024}\natexlab{b}.
\newblock \showarticletitle{Lamp: When large language models meet
  personalization}. In \bibinfo{booktitle}{\emph{Proceedings of the 62nd Annual
  Meeting of the Association for Computational Linguistics (Volume 1: Long
  Papers)}}. \bibinfo{pages}{7370--7392}.
\newblock


\bibitem[Salemi and Zamani(2025)]%
        {salemi2024ragvsft}
\bibfield{author}{\bibinfo{person}{Alireza Salemi} {and} \bibinfo{person}{Hamed
  Zamani}.} \bibinfo{year}{2025}\natexlab{}.
\newblock \showarticletitle{Comparing Retrieval-Augmentation and
  Parameter-Efficient Fine-Tuning for Privacy-Preserving Personalization of
  Large Language Models}. In \bibinfo{booktitle}{\emph{Proceedings of the 2025
  International ACM SIGIR Conference on Innovative Concepts and Theories in
  Information Retrieval (ICTIR)}} (Padua, Italy) \emph{(\bibinfo{series}{ICTIR
  '25})}. \bibinfo{publisher}{Association for Computing Machinery},
  \bibinfo{address}{New York, NY, USA}, \bibinfo{pages}{286–296}.
\newblock
\showISBNx{9798400718618}
\href{https://doi.org/10.1145/3731120.3744595}{doi:\nolinkurl{10.1145/3731120.3744595}}


\bibitem[Song and Tang(2025)]%
        {mobilemebpsong25}
\bibfield{author}{\bibinfo{person}{Congzheng Song} {and} \bibinfo{person}{Xinyu
  Tang}.} \bibinfo{year}{2025}\natexlab{}.
\newblock \showarticletitle{Memory-Efficient Backpropagation for Fine-Tuning
  LLMs on Resource-Constrained Mobile Devices}. In
  \bibinfo{booktitle}{\emph{Proceedings of the 2025 Conference on Empirical
  Methods in Natural Language Processing, {EMNLP} 2025 - Industry Track,
  Suzhou, China, November 4-9, 2025}},
  \bibfield{editor}{\bibinfo{person}{Saloni Potdar},
  \bibinfo{person}{Lina~Maria Rojas{-}Barahona}, {and}
  \bibinfo{person}{S{\'{e}}bastien Montella}} (Eds.).
  \bibinfo{publisher}{Association for Computational Linguistics},
  \bibinfo{pages}{766--774}.
\newblock
\href{https://doi.org/10.18653/V1/2025.EMNLP-INDUSTRY.52}{doi:\nolinkurl{10.18653/V1/2025.EMNLP-INDUSTRY.52}}


\bibitem[Sung et~al\mbox{.}(2022)]%
        {sung2022lst}
\bibfield{author}{\bibinfo{person}{Yi-Lin Sung}, \bibinfo{person}{Jaemin Cho},
  {and} \bibinfo{person}{Mohit Bansal}.} \bibinfo{year}{2022}\natexlab{}.
\newblock \showarticletitle{{LST}: Ladder Side-Tuning for Parameter and Memory
  Efficient Transfer Learning}. In \bibinfo{booktitle}{\emph{Advances in Neural
  Information Processing Systems 35 (NeurIPS 2022)}}.
\newblock
\showeprint[arxiv]{2206.06522}~[cs.LG]
\urldef\tempurl%
\url{https://proceedings.neurips.cc/paper_files/paper/2022/hash/54801e196796134a2b0ae5e8adef502f-Abstract-Conference.html}
\showURL{%
\tempurl}


\bibitem[Tam et~al\mbox{.}(2026)]%
        {fblayout_tam26}
\bibfield{author}{\bibinfo{person}{Kahou Tam}, \bibinfo{person}{Wei Niu},
  \bibinfo{person}{Yu Bao}, \bibinfo{person}{Xiaomin Ouyang},
  \bibinfo{person}{ChengZhong Xu}, {and} \bibinfo{person}{Li Li}.}
  \bibinfo{year}{2026}\natexlab{}.
\newblock \showarticletitle{FBLayout: Optimizing Memory Layout for Efficient
  {LLM} Finetuning on Mobile GPUs}. In \bibinfo{booktitle}{\emph{Proceedings of
  the 24th Annual International Conference on Mobile Systems, Applications and
  Services, MobiSys 2026, University of Cambridge, Cambridge, United Kingdom,
  June 21-25, 2026}}, \bibfield{editor}{\bibinfo{person}{Cecilia Mascolo} {and}
  \bibinfo{person}{Fahim Kawsar}} (Eds.). \bibinfo{publisher}{{ACM}},
  \bibinfo{pages}{432--447}.
\newblock
\href{https://doi.org/10.1145/3745756.3809214}{doi:\nolinkurl{10.1145/3745756.3809214}}


\bibitem[Tan et~al\mbox{.}(2024)]%
        {oppu2024}
\bibfield{author}{\bibinfo{person}{Zhaoxuan Tan}, \bibinfo{person}{Qingkai
  Zeng}, \bibinfo{person}{Yijun Tian}, \bibinfo{person}{Zheyuan Liu},
  \bibinfo{person}{Bing Yin}, {and} \bibinfo{person}{Meng Jiang}.}
  \bibinfo{year}{2024}\natexlab{}.
\newblock \showarticletitle{Democratizing Large Language Models via
  Personalized Parameter-Efficient Fine-tuning}. In
  \bibinfo{booktitle}{\emph{Proceedings of the 2024 Conference on Empirical
  Methods in Natural Language Processing}},
  \bibfield{editor}{\bibinfo{person}{Yaser Al-Onaizan}, \bibinfo{person}{Mohit
  Bansal}, {and} \bibinfo{person}{Yun-Nung Chen}} (Eds.).
  \bibinfo{publisher}{Association for Computational Linguistics},
  \bibinfo{address}{Miami, Florida, USA}, \bibinfo{pages}{6476--6491}.
\newblock
\href{https://doi.org/10.18653/v1/2024.emnlp-main.372}{doi:\nolinkurl{10.18653/v1/2024.emnlp-main.372}}


\bibitem[{Tencent}(2017)]%
        {ncnn}
\bibfield{author}{\bibinfo{person}{{Tencent}}.}
  \bibinfo{year}{2017}\natexlab{}.
\newblock \bibinfo{title}{{ncnn}: A High-Performance Neural Network Inference
  Framework Optimized for the Mobile Platform}.
\newblock \bibinfo{howpublished}{\url{https://github.com/Tencent/ncnn}}.
\newblock


\bibitem[Tenison et~al\mbox{.}(2026)]%
        {tenison2026lars}
\bibfield{author}{\bibinfo{person}{Irene Tenison}, \bibinfo{person}{Stella
  Ahn}, \bibinfo{person}{Miriam Kim}, \bibinfo{person}{Ebtisam Alshehri}, {and}
  \bibinfo{person}{Lalana Kagal}.} \bibinfo{year}{2026}\natexlab{}.
\newblock \bibinfo{title}{Parameter Efficiency Is Not Memory Efficiency:
  Rethinking Fine-Tuning for On-Device {LLM} Adaptation}.
\newblock
\showeprint[arxiv]{2604.22783}~[cs.LG]
\urldef\tempurl%
\url{https://arxiv.org/abs/2604.22783}
\showURL{%
\tempurl}


\bibitem[Tummalapalli et~al\mbox{.}(2026)]%
        {tummalapalli2026edgesustained}
\bibfield{author}{\bibinfo{person}{Pranay Tummalapalli}, \bibinfo{person}{Sahil
  Arayakandy}, \bibinfo{person}{Ritam Pal}, {and} \bibinfo{person}{Kautuk
  Kundan}.} \bibinfo{year}{2026}\natexlab{}.
\newblock \bibinfo{title}{{LLM} Inference at the Edge: Mobile, {NPU}, and {GPU}
  Performance Efficiency Trade-offs Under Sustained Load}.
\newblock
\showeprint[arxiv]{2603.23640}~[cs.DC]
\urldef\tempurl%
\url{https://arxiv.org/abs/2603.23640}
\showURL{%
\tempurl}


\bibitem[Vepakomma et~al\mbox{.}(2018)]%
        {splitlearningvepakomma2018}
\bibfield{author}{\bibinfo{person}{Praneeth Vepakomma},
  \bibinfo{person}{Otkrist Gupta}, \bibinfo{person}{Tristan Swedish}, {and}
  \bibinfo{person}{Ramesh Raskar}.} \bibinfo{year}{2018}\natexlab{}.
\newblock \showarticletitle{Split learning for health: Distributed deep
  learning without sharing raw patient data}.
\newblock \bibinfo{journal}{\emph{CoRR}}  \bibinfo{volume}{abs/1812.00564}
  (\bibinfo{year}{2018}).
\newblock
\showeprint[arXiv]{1812.00564}
\urldef\tempurl%
\url{http://arxiv.org/abs/1812.00564}
\showURL{%
\tempurl}


\bibitem[Wang et~al\mbox{.}(2025)]%
        {wang2025xpert}
\bibfield{author}{\bibinfo{person}{Haoming Wang}, \bibinfo{person}{Boyuan
  Yang}, \bibinfo{person}{Xiangyu Yin}, {and} \bibinfo{person}{Wei Gao}.}
  \bibinfo{year}{2025}\natexlab{}.
\newblock \showarticletitle{Never Start from Scratch: Expediting On-Device
  {LLM} Personalization via Explainable Model Selection}. In
  \bibinfo{booktitle}{\emph{Proceedings of the 23rd Annual International
  Conference on Mobile Systems, Applications and Services (MobiSys '25)}}.
  \bibinfo{pages}{154--168}.
\newblock
\href{https://doi.org/10.1145/3711875.3729132}{doi:\nolinkurl{10.1145/3711875.3729132}}


\bibitem[Wang et~al\mbox{.}(2022)]%
        {wang2022melon}
\bibfield{author}{\bibinfo{person}{Qipeng Wang}, \bibinfo{person}{Mengwei Xu},
  \bibinfo{person}{Chao Jin}, \bibinfo{person}{Xinran Dong},
  \bibinfo{person}{Jinliang Yuan}, \bibinfo{person}{Xin Jin},
  \bibinfo{person}{Gang Huang}, \bibinfo{person}{Yunxin Liu}, {and}
  \bibinfo{person}{Xuanzhe Liu}.} \bibinfo{year}{2022}\natexlab{}.
\newblock \showarticletitle{{Melon}: breaking the memory wall for
  resource-efficient on-device machine learning}. In
  \bibinfo{booktitle}{\emph{Proceedings of the 20th Annual International
  Conference on Mobile Systems, Applications and Services (MobiSys '22)}}.
  \bibinfo{publisher}{{ACM}}, \bibinfo{pages}{450--463}.
\newblock
\href{https://doi.org/10.1145/3498361.3538928}{doi:\nolinkurl{10.1145/3498361.3538928}}


\bibitem[Wang et~al\mbox{.}(2026b)]%
        {wang2026tzllm}
\bibfield{author}{\bibinfo{person}{Xunjie Wang}, \bibinfo{person}{Jiacheng
  Shi}, \bibinfo{person}{Zihan Zhao}, \bibinfo{person}{Yang Yu},
  \bibinfo{person}{Zhichao Hua}, {and} \bibinfo{person}{Jinyu Gu}.}
  \bibinfo{year}{2026}\natexlab{b}.
\newblock \showarticletitle{{TZ-LLM}: Protecting On-Device Large Language
  Models with {Arm} {TrustZone}}. In \bibinfo{booktitle}{\emph{Proceedings of
  the 21st European Conference on Computer Systems (EuroSys '26)}}.
  \bibinfo{publisher}{{ACM}}, \bibinfo{pages}{657--674}.
\newblock
\href{https://doi.org/10.1145/3767295.3769334}{doi:\nolinkurl{10.1145/3767295.3769334}}


\bibitem[Wang et~al\mbox{.}(2023)]%
        {wang2023egeria}
\bibfield{author}{\bibinfo{person}{Yiding Wang}, \bibinfo{person}{Decang Sun},
  \bibinfo{person}{Kai Chen}, \bibinfo{person}{Fan Lai}, {and}
  \bibinfo{person}{Mosharaf Chowdhury}.} \bibinfo{year}{2023}\natexlab{}.
\newblock \showarticletitle{Egeria: Efficient {DNN} Training with
  Knowledge-Guided Layer Freezing}. In \bibinfo{booktitle}{\emph{Proceedings of
  the Eighteenth European Conference on Computer Systems (EuroSys '23)}}.
  \bibinfo{publisher}{{ACM}}, \bibinfo{pages}{851--866}.
\newblock
\href{https://doi.org/10.1145/3552326.3587451}{doi:\nolinkurl{10.1145/3552326.3587451}}


\bibitem[Wang et~al\mbox{.}(2026a)]%
        {wang2026edgetune}
\bibfield{author}{\bibinfo{person}{Zhenyu Wang}, \bibinfo{person}{Rana
  Muhammad~Shahroz Khan}, \bibinfo{person}{Tianlong Chen}, {and}
  \bibinfo{person}{Shahriar Nirjon}.} \bibinfo{year}{2026}\natexlab{a}.
\newblock \showarticletitle{{EdgeTune}: Efficient On-Device {LLM}
  Personalization at the Edge}. In \bibinfo{booktitle}{\emph{Proceedings of the
  2026 ACM/IEEE International Conference on Embedded Artificial Intelligence
  and Sensing Systems (SenSys '26)}}. \bibinfo{publisher}{{ACM}},
  \bibinfo{pages}{421--437}.
\newblock
\href{https://doi.org/10.1145/3774906.3802769}{doi:\nolinkurl{10.1145/3774906.3802769}}


\bibitem[Xu et~al\mbox{.}(2022)]%
        {xu2022mandheling}
\bibfield{author}{\bibinfo{person}{Daliang Xu}, \bibinfo{person}{Mengwei Xu},
  \bibinfo{person}{Qipeng Wang}, \bibinfo{person}{Shangguang Wang},
  \bibinfo{person}{Yun Ma}, \bibinfo{person}{Kang Huang}, \bibinfo{person}{Gang
  Huang}, \bibinfo{person}{Xin Jin}, {and} \bibinfo{person}{Xuanzhe Liu}.}
  \bibinfo{year}{2022}\natexlab{}.
\newblock \showarticletitle{{Mandheling}: Mixed-Precision On-Device {DNN}
  Training with {DSP} Offloading}. In \bibinfo{booktitle}{\emph{Proceedings of
  the 28th Annual International Conference on Mobile Computing and Networking
  (MobiCom '22)}}. \bibinfo{publisher}{{ACM}}, \bibinfo{pages}{214--227}.
\newblock
\href{https://doi.org/10.1145/3495243.3560545}{doi:\nolinkurl{10.1145/3495243.3560545}}


\bibitem[Yang and Xia(2026)]%
        {yang2026pelm}
\bibfield{author}{\bibinfo{person}{Weisi Yang} {and} \bibinfo{person}{Stephen
  Xia}.} \bibinfo{year}{2026}\natexlab{}.
\newblock \showarticletitle{{PELM}: Power Efficient On-Device {LLM} Inference
  with Speculative Decoding and Dynamic Voltage Frequency Scaling}. In
  \bibinfo{booktitle}{\emph{Proceedings of the 2026 ACM/IEEE International
  Conference on Embedded Artificial Intelligence and Sensing Systems (SenSys
  '26)}}. \bibinfo{publisher}{{ACM}}.
\newblock
\newblock
\shownote{arXiv:2609.09662}.


\bibitem[Yang et~al\mbox{.}(2025)]%
        {yang2025paemobillm}
\bibfield{author}{\bibinfo{person}{Xingke Yang}, \bibinfo{person}{Liang Li},
  \bibinfo{person}{Zhiyi Wan}, \bibinfo{person}{Sicong Li},
  \bibinfo{person}{Xiaoqi Qi}, \bibinfo{person}{Jiang Liu},
  \bibinfo{person}{Tomoaki Ohtsuki}, \bibinfo{person}{Xin Fu}, {and}
  \bibinfo{person}{Miao Pan}.} \bibinfo{year}{2025}\natexlab{}.
\newblock \bibinfo{title}{{PAE MobiLLM}: Privacy-Aware and Efficient {LLM}
  Fine-Tuning on the Mobile Device via Additive Side-Tuning}.
\newblock
\showeprint[arxiv]{2507.01216}~[cs.LG]
\urldef\tempurl%
\url{https://arxiv.org/abs/2507.01216}
\showURL{%
\tempurl}


\bibitem[Zhang et~al\mbox{.}(2025a)]%
        {zhang2025dvfs}
\bibfield{author}{\bibinfo{person}{Zongpu Zhang}, \bibinfo{person}{Pranab
  Dash}, \bibinfo{person}{Y.~Charlie Hu}, \bibinfo{person}{Qiang Xu},
  \bibinfo{person}{Jian Li}, {and} \bibinfo{person}{Haibing Guan}.}
  \bibinfo{year}{2025}\natexlab{a}.
\newblock \bibinfo{title}{Dissecting the Impact of Mobile {DVFS} Governors on
  {LLM} Inference Performance and Energy Efficiency}.
\newblock
\showeprint[arxiv]{2507.02135}~[cs.OS]
\urldef\tempurl%
\url{https://arxiv.org/abs/2507.02135}
\showURL{%
\tempurl}


\bibitem[Zhang et~al\mbox{.}(2025b)]%
        {zhang2025personalizationlargelanguagemodels}
\bibfield{author}{\bibinfo{person}{Zhehao Zhang}, \bibinfo{person}{Ryan~A.
  Rossi}, \bibinfo{person}{Branislav Kveton}, \bibinfo{person}{Yijia Shao},
  \bibinfo{person}{Diyi Yang}, \bibinfo{person}{Hamed Zamani},
  \bibinfo{person}{Franck Dernoncourt}, \bibinfo{person}{Joe Barrow},
  \bibinfo{person}{Tong Yu}, \bibinfo{person}{Sungchul Kim},
  \bibinfo{person}{Ruiyi Zhang}, \bibinfo{person}{Jiuxiang Gu},
  \bibinfo{person}{Tyler Derr}, \bibinfo{person}{Hongjie Chen},
  \bibinfo{person}{Junda Wu}, \bibinfo{person}{Xiang Chen},
  \bibinfo{person}{Zichao Wang}, \bibinfo{person}{Subrata Mitra},
  \bibinfo{person}{Nedim Lipka}, \bibinfo{person}{Nesreen Ahmed}, {and}
  \bibinfo{person}{Yu Wang}.} \bibinfo{year}{2025}\natexlab{b}.
\newblock \showarticletitle{Personalization of Large Language Models: A
  Survey}.
\newblock  (\bibinfo{year}{2025}).
\newblock
\showeprint[arxiv]{2411.00027}~[cs.CL]
\urldef\tempurl%
\url{https://arxiv.org/abs/2411.00027}
\showURL{%
\tempurl}


\bibitem[Zhong et~al\mbox{.}(2025)]%
        {mobicdcda2025}
\bibfield{author}{\bibinfo{person}{Zhaofeng Zhong}, \bibinfo{person}{Wei Yuan},
  \bibinfo{person}{Liang Qu}, \bibinfo{person}{Tong Chen}, \bibinfo{person}{Hao
  Wang}, \bibinfo{person}{Xiangyu Zhao}, {and} \bibinfo{person}{Hongzhi Yin}.}
  \bibinfo{year}{2025}\natexlab{}.
\newblock \showarticletitle{Towards On-Device Personalization: Cloud-device
  Collaborative Data Augmentation for Efficient On-device Language Model}.
\newblock \bibinfo{journal}{\emph{arXiv preprint arXiv:2508.21313}}
  (\bibinfo{year}{2025}).
\newblock


\bibitem[Zhu et~al\mbox{.}(2023)]%
        {zhu2023pockengine}
\bibfield{author}{\bibinfo{person}{Ligeng Zhu}, \bibinfo{person}{Lanxiang Hu},
  \bibinfo{person}{Ji Lin}, \bibinfo{person}{Wei-Ming Chen},
  \bibinfo{person}{Wei-Chen Wang}, \bibinfo{person}{Chuang Gan}, {and}
  \bibinfo{person}{Song Han}.} \bibinfo{year}{2023}\natexlab{}.
\newblock \showarticletitle{{PockEngine}: Sparse and Efficient Fine-tuning in a
  Pocket}. In \bibinfo{booktitle}{\emph{Proceedings of the 56th Annual IEEE/ACM
  International Symposium on Microarchitecture (MICRO '23)}}.
  \bibinfo{publisher}{{ACM}}, \bibinfo{pages}{1381--1394}.
\newblock
\href{https://doi.org/10.1145/3613424.3614307}{doi:\nolinkurl{10.1145/3613424.3614307}}


\bibitem[Zhu et~al\mbox{.}(2019)]%
        {zhu2019dlg}
\bibfield{author}{\bibinfo{person}{Ligeng Zhu}, \bibinfo{person}{Zhijian Liu},
  {and} \bibinfo{person}{Song Han}.} \bibinfo{year}{2019}\natexlab{}.
\newblock \showarticletitle{Deep leakage from gradients}.
\newblock \bibinfo{journal}{\emph{Advances in neural information processing
  systems}}  \bibinfo{volume}{32} (\bibinfo{year}{2019}).
\newblock


\bibitem[Zhu et~al\mbox{.}(2026)]%
        {zhu2026lorafusion}
\bibfield{author}{\bibinfo{person}{Zhanda Zhu}, \bibinfo{person}{Qidong Su},
  \bibinfo{person}{Yaoyao Ding}, \bibinfo{person}{Kevin Song},
  \bibinfo{person}{Shang Wang}, {and} \bibinfo{person}{Gennady Pekhimenko}.}
  \bibinfo{year}{2026}\natexlab{}.
\newblock \showarticletitle{{LoRAFusion}: Efficient {LoRA} Fine-Tuning for
  {LLMs}}. In \bibinfo{booktitle}{\emph{Proceedings of the 21st European
  Conference on Computer Systems (EuroSys '26)}}. \bibinfo{publisher}{{ACM}},
  \bibinfo{pages}{586--604}.
\newblock
\href{https://doi.org/10.1145/3767295.3769331}{doi:\nolinkurl{10.1145/3767295.3769331}}


\bibitem[Zou et~al\mbox{.}(2026)]%
        {zou2026enerinfer}
\bibfield{author}{\bibinfo{person}{Bohua Zou}, \bibinfo{person}{Nian Liu},
  \bibinfo{person}{Binqi Sun}, \bibinfo{person}{Matteo Mascherin},
  \bibinfo{person}{Debayan Roy}, \bibinfo{person}{Yutao Liu},
  \bibinfo{person}{Yu Peng}, \bibinfo{person}{Ning Jia}, {and}
  \bibinfo{person}{Haibo Chen}.} \bibinfo{year}{2026}\natexlab{}.
\newblock \bibinfo{title}{{EnerInfer}: Energy-Aware On-Device {LLM} Inference}.
\newblock
\showeprint[arxiv]{2606.23001}~[cs.SE]
\urldef\tempurl%
\url{https://arxiv.org/abs/2606.23001}
\showURL{%
\tempurl}


\end{thebibliography}
